\documentclass[aps,pra,showpacs,twocolumn,amssymb,floats,epsfig]{revtex4}

\usepackage{graphicx}
\usepackage{epsfig}
\usepackage{amsfonts}
\usepackage{amsmath}
\usepackage{amssymb}
\usepackage{color}

\newcommand\ba{\begin{eqnarray}}
\newcommand\ea{\end{eqnarray}}
\newcommand\be{\begin{equation}}
\newcommand\ee{\end{equation}}
\newcommand\bi{\bibitem}
\newcommand{\ct}{\cite}
\newcommand{\bra}[1]{\langle #1|}
\newcommand{\ket}[1]{|#1\rangle}

\def\non{\nonumber}

\def\de{\delta}

\def\la{\lambda}

\begin{document}

\title{Generation of concurrence   between two qubits locally coupled to a one dimensional spin chain }
\author{ Tanay Nag}
\author{Amit Dutta}
\affiliation{\small{
Department of Physics, Indian Institute of Technology, Kanpur 208 016, 
India}}
\begin{abstract}
 We consider a generalized central spin model, consisting of two central qubits and an environmental 
 spin chain (with periodic boundary condition) to which these central qubits are locally and weakly connected either at the same site or at
 two different sites separated by a distance $d$. Our purpose is to study the subsequent temporal generation of entanglement,
 quantified by concurrence, when  initially  the qubits are in an unentangled state. In the equilibrium situation, we show that the
 concurrence survives for a larger value of $d$  when the environmental spin chain is critical. Importantly, a common feature observed 
both  in the equilibrium and the non-equilibrium situations while the latter is
created by a sudden but global change of the environmental transverse field, is
that the two qubits become maximally entangled for the critical quenching. Following a non-equilibrium 
evolution of the spin chain, our study for $d\neq 0$, indicates that there exists a threshold time above which 
concurrence attains a finite value. Additionally, we show that the number of independent decohering channels (DCs) is determined by $d$
as well as the  local difference of the transverse field of the two underlying Hamiltonians governing the time evolution.
The qualitatively similar behavior displayed by the concurrence for critical and off-critical quenches, as reported here, is characterized
by analyzing the non-equilibrium evolution of  these  channels.  The concurrence is maximum  when the {decoherence factor or the echo associated with the}
most rapidly DC decays to zero; on the contrary, the condition when the concurrence vanishes is determined non-trivially
 by the associated decay of one of the  intermediate DCs. Analyzing the reduced density of a single qubit, 
 we also explain the  observation that  the dephasing rate is always slower than the unentanglement rate. We further show that the maximally and minimally
decohering channels show singular behavior which can be explained by invoking upon a quasi-particle picture.
 
 \end{abstract}

\maketitle

\section{introduction}
\label{intro}
The notion of entanglement, that emerged from  the pioneering work of Einstein, Podolsky and Rosen \ct{einstein35}, is a key concept
 of quantum computation and quantum information theories \ct{werner89,bennett96,nielsen00,vedral07}. Given the recent interest in the 
 studies of quantum correlations in the context of quantum critical systems \ct{chakrabarti96,sachdev99,polkovnikov11,dutta15}, there have been numerous efforts 
 directed to
 understanding the connection between quantum information and  quantum phase transitions (QPTs) \ct{osterloh02,osborne02,vidal03,wu04,gu04,chen06,amico08}.
 Entanglement is usually quantified through  two  quantum information theoretic measures: (i) concurrence \ct{wootters01,horodecki01,horodecki09},
 a separability based approach to measure the quantum correlation and (ii) quantum discord \ct{ollivier01,luo08,sarandy09},
 a measurement based approach for estimating the non-classical correlations present in a bipartite 
 system.  There have also  been numerous studies on the entanglement entropy which is another important tool to probe the entanglement between two blocks of a composite
 system obtained by measuring the von Neumann entropy associated with  the reduced density matrix of one of the blocks \ct{calabrese05}.

 It is now established that the effect of  quantum criticality gets imprinted in the behavior of the ground state correlation which
 becomes maximum at the quantum critical point (QCP);  for example, the concurrence can  detect as well as characterize a QPT \ct{osterloh02,osborne02}.
  On the other hand, the entanglement, arising due to the  interaction between the system and its environment,  leads to  
  decoherence  \ct{joos03}. There exists a plethora of the studies investigating the 
 effects  induced by the environment on  the quantum information processing \ct{plenio99}; simultaneously, the dynamical control of the 
 decoherence is also being investigated extensively \ct{viola98,rossini08}.

  The central spin model (CSM), consisting of a single qubit (spin-1/2) globally coupled to an environmental spin chain, is  an important 
  prototypical model to study the Loschmidt echo (LE), also known as the decoherence factor (DF) characterising the decoherence of the qubit; this has
  been studied   for both equilibrium
 \ct{cucchietti05,quan06,sharma12} and
  non-equilibrium \ct{damski11a,nag12a,nag12b,roy13,patel13,mukherjee12,sharma14,suzuki15,nag15} time evolution. 
Moreover, the concurrence \ct{sengupta09} and the quantum discord \ct{nag11} have been  shown to satisfy 
 the universal scaling law as predicted by the Kibble-Zurek argument \ct{zurek05,polkovnikov05} when a parameter of the environmental Hamiltonian is driven linearly across a QCP.
  Additionally, a generalized central spin model (GCSM)   where  
  two spins  are  globally coupled to a  environmental spin chain, with a periodic boundary condition (PBC),
  is also  studied for probing the concurrence and quantum  discord generated between the qubits when the composite system evolves in time \ct{yuan07,sun07,liu10}; the concurrence generation is found to be maximum
 for the critical spin chain \ct{yi06,ai08}.  In connection to the experimental studies,
  a QPT has already been observed with ultracold atoms in an optical lattice \ct{greiner02}. A possible realization of a one dimensional 
  $XY$ chain has also been proposed \ct{duan03}. Furthermore, using NMR quantum simulator, it has been experimentally confirmed  that the LE shows a 
  dip at the QCP of a finite antiferromagnetic Ising spin chain, thereby establishing it an ideal detector of a QCP  \ct{zhang09}.

Recently, there have been  investigations \ct{cormick08,wendenbaum14} of the unentanglement (i.e., the decoherence) between two distant qubits initially entangled and connected to
   two different sites of the spin chain  that  evolves in time.
      The state transfer quality between two external qubits  of a spin chain has also been  investigated by analyzing  the entanglement between them \ct{hartmann06}.  Given the previous studies,  we address  the reverse question.
   Is there a temporal  generation of entanglement between a pair of qubits,
    initially prepared in an unentangled state, connected at the same site and also two different sites (separated by a distance $d$) of
 the environmental spin chain? To address this particular issue, we  consider  two situations: (i) when the spin chain (chosen to be a one-dimensional transverse Ising chain)  evolves temporally in time following the local coupling of the qubits, referred to
 as the equilibrium situation.
 (ii) There is an additional sudden global quench of the transverse field of the environmental Hamiltonian  in addition to the local coupling, referred to as the non-equilibrium 
 situation. To the best of our knowledge, ours is the first work to investigate the  generation of  the concurrence 
in the  non-equilibrium situation using a GCSM with a local coupling.

 We briefly summarize our main results at the outset: firstly, we are working  in the  weak coupling limit considering a PBC for the environmental spin chain. Our observation for the equilibrium situation is that the entanglement
 generation is of very small magnitude  although, in the vicinity of the QCP of the environment, the concurrence becomes maximum and remains
 finite even when the two qubits are separated by a large distance.
 In the non-equilibrium situation, our investigation suggests that the concurrence, of much higher magnitude than the equilibrium case, can be induced by the 
 global and sudden quench of the transverse field;  this concurrence  eventually 
 decays with time. We explain this generic behavior of the concurrence by analyzing the echoes associated with different decohering channels (DCs) 
 with time; we observe that the number of independent channels 
 is dictated by the  separation $d$  and local difference of the transverse field of the two underlying Hamiltonian governing the time evolution. 
 The decay of most rapidly DC is responsible for maximum amount of entanglement while the decay of the concurrence is non-trivially 
 related to the decay of one of the intermediate decaying channel.  For a finite separation between the qubits, we establish  that the concurrence attains a non-zero value after a threshold
 time; this is attributed to the behavior of different DC up to the threshold time.
 Additionally, we show that the respective dephasing rate associated with each qubit is always slower than the unentanglement rate
 between the qubits. Furthermore, we characterize the
 distance dependent behavior of different DCs following a critical quench by making resort to a  quasi-particles picture.
 
The paper is organized in the following manner. In Sec.\ref{model}, we introduce the CGCM consisting of two qubit 
locally connected to two sites of an Ising chain with a transverse field. In parallel, we define the concurrence derived from the $4 \times 4$ reduced
density matrix  of the two qubits obtained by tracing out the environmental degrees of freedom; this density matrix contains  different LEs corresponding to different DCs associated with the environmental
evolution. In Sec.\ref{result1} the results obtained
for the equilibrium situation are presented while in Sec.\ref{result2}, we discuss the non-equilibrium behavior of the concurrence. We analyze the behavior of 
concurrence observed in equilibrium as well as non-equilibrium cases investigating  the temporal evolution of different decohering 
channels. Finally, we make concluding remarks in Sec.\ref{conclude}.

\section{model}
\label{model}

\begin{figure}[ht]
\begin{center}
\includegraphics[height=5.0cm,width=8.1cm]{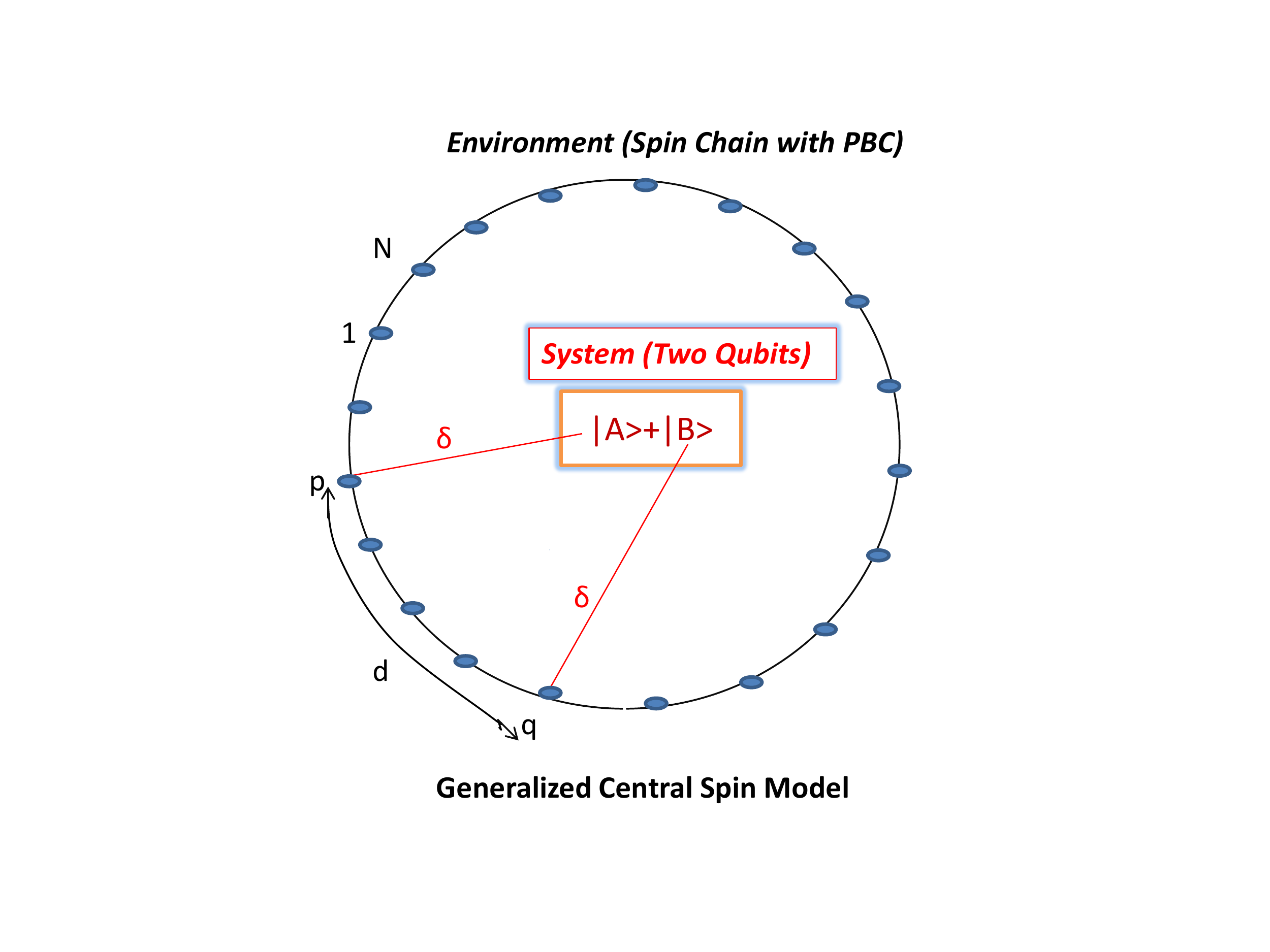}
\end{center}
\caption{(Color online) Schematic diagram shows the generalized central spin model where two qubits $A$ and $B$ are locally connected to two 
different environmental sites $p$ and $q$ separated from each other by a distance $d$ with a coupling strength $\delta$. 
We consider a periodic boundary condition for the environmental spin chain with $N$ number of spins.
 } \label{fig10} \end{figure}

We consider a GCSM  in which  two non-interacting qubits are connected by a local interaction  to an environmental spin chain, chosen to be a one-dimensional 
ferromagnetic transverse Ising spin model, in such a way that the local  transverse field of the environmental spin chain gets modified.
The composite system, thus,  is a generalization of the central spin model \cite{quan06},
in which a single spin-$1/2$ particle (qubit) is globally connected to all the spins of the 
environmental spin chain with an interaction Hamiltonian; the schematic diagram of the GCSM is shown in Fig.~(\ref{fig10}). The combined Hamiltonian  $H_{\rm T}$, 
comprising of an environmental transverse Ising Hamiltonian $H_{\rm E}$ \ct{dutta15} with $N$ number of spins and interaction
Hamiltonian $H_{\rm SE}$ of two qubits, is given by
\be
H_T=H_{SE}+H_E.
\label{eq:model}
\ee
Here,
\be
H_E~=~-J\sum_{n}\sigma_n^x\sigma_{n+1}^x~-~\la\sum_{n}\sigma_n^z,
\label{eq:model_e}
\ee
where $\sigma$'s are the usual Pauli matrices. $\la$ and $J$ (set equal to unity below) are the  transverse magnetic field and ferromagnetic cooperative interactions, respectively.
We consider a PBC $\sigma^i_{N+1}=\sigma^i_1$. 
The interaction Hamiltonian for two qubits $A$ and $B$ connected at 
different sites of the environment is given by 
\be
H_{SE}=-\delta(\ket{\uparrow}\bra{\uparrow}_A\otimes \sigma^z_p
+\ket{\uparrow}\bra{\uparrow}_B\otimes \sigma^z_q);
\label{eq:model_s}
\ee
here, $\ket{\uparrow}_{A,B}$ is an eigenstate of $\sigma^z_{A,B}$ satisfying
$\sigma^z_{A,B}\ket{\uparrow}_{A,B}=\ket{\uparrow}_{A,B}$ while$\sigma^z_{p,q}$ denote the 
environmental spin at $p$ and $q$-th site, respectively; these sites are separated by a distance $d$.
$\delta$ is the 
coupling strength and   we shall work in the limit $\de\to0$. Clearly,  the interaction Hamiltonian \eqref{eq:model_s} suggests interaction with the qubits modifies the 
local transverse field of the environment.

In order to study the generation of concurrence between two qubits, we take a completely 
unentangled  (direct product)  initial state given by:
\be
\ket{\phi}_{AB}={1 \over 2}(\ket{\uparrow}_A+\ket{\downarrow}_A)\otimes
(\ket{\uparrow}_B+\ket{\downarrow}_B).
\label{eq:qubit}
\ee
The initial state for the composite system is then given by
$\ket{\psi(\la_i,t=0)}=\ket{\phi}_{AB}\otimes\ket{\eta(\la_i,t=0)}$, where $\ket{\eta(\la_i,t=0)}$
is the initial ground state of the environmental Hamiltonian $H_E$ given in Eq.~(\ref{eq:model_e}).

Focusing on the non-equilibrium situation, we consider a sudden quenching of the transverse 
field which is instantaneously changed from an initial value $\la_i$ to a final value $\la_f$ and study the
subsequent temporal   evolution of the composite system.  We note that the equilibrium situation corresponds to $\la_f =\la_i$.
 Depending upon the state of the qubits, the interaction Hamiltonian leads to four channels 
of evolution for the environment. The channel Hamiltonians $H_{\alpha \beta}$ with $\la_f$ governing the 
dynamics are given by 
\ba
H_{\downarrow \downarrow}(\la_f)&=& H_E(\la_f), \nonumber \\
H_{\uparrow \uparrow}(\la_f)&=& H_E(\la_f)-\delta(\sigma^z_p+\sigma^z_q),\nonumber \\
H_{\downarrow \uparrow}(\la_f)&=& H_E(\la_f)-\delta \sigma^z_q, \nonumber \\
H_{\uparrow \downarrow}(\la_f)&=& H_E(\la_f)-\delta \sigma^z_p.
\label{eq:channel}
\ea

The time evolved state of the composite system is given by
\ba
\ket{\psi(t)}&=& {1 \over 2} \biggl( \ket{\uparrow \uparrow} \otimes \ket{\eta_{\uparrow \uparrow}(t)} +
 \ket{\downarrow \downarrow} \otimes \ket{\eta_{\downarrow \downarrow}(t)}   \nonumber \\
&+&  \ket{\uparrow \downarrow} \otimes \ket{\eta_{\uparrow \downarrow}(t)} +
 \ket{\downarrow \uparrow} \otimes \ket{\eta_{\downarrow \uparrow}(t)} \biggr),
\label{eq:state_t}
\ea
where $\ket{\alpha\beta}$ represents the state for two qubits and environmental evolved state $\ket{\eta_{\alpha\beta}(t)}$
is given by 
\be
\ket{\eta_{\alpha \beta}(t)}=e^{-iH_{\alpha\beta}t}\ket{\eta(\la_i,t=0)},
\label{eq:state_e}
\ee
where $\la_i$ is the initial homogeneous transverse field same for all sites i.e., $\la_n=\la_i$.

One can construct the reduced density matrix of the qubits by tracing out the environmental
degrees of freedom from the composite density matrix constructed from 
$\ket{\psi(t)}$. The reduced density matrix for the two qubits system in the basis 
{$\{\ket{\uparrow\uparrow},\ket{\uparrow\downarrow},\ket{\downarrow\uparrow},\ket{\downarrow\downarrow}\}$} is given by
\be \rho_s(t) = \frac {1}{4} \left[ \begin{array}{cccc} 
1 & d_{\uparrow \uparrow, \uparrow \downarrow} & d_{\uparrow \uparrow, \downarrow \uparrow}& d_{\uparrow \uparrow, \downarrow \downarrow} \\
d^*_{\uparrow \uparrow, \uparrow \downarrow} & 1 & d_{\uparrow \downarrow, \downarrow \uparrow}& d_{\uparrow \downarrow, \downarrow \downarrow} \\
d^*_{\uparrow \uparrow, \downarrow \uparrow} & d^*_{\uparrow \downarrow, \downarrow \uparrow}& 1 & d_{\downarrow \uparrow, \downarrow \downarrow} \\
d^*_{\uparrow \uparrow, \downarrow \downarrow} & d^*_{\uparrow \downarrow, \downarrow \downarrow}& d^*_{\downarrow \uparrow, \downarrow \downarrow} & 1 
\end{array} \right], \label{eq:rho_s} \ee
where $d_{\alpha\beta,\gamma\lambda}=\langle \eta_{\alpha\beta}(t)|\eta_{\gamma\lambda}(t)\rangle$.
The DFs or the echoes corresponding to different channels are  $D_{\alpha\beta,\gamma\lambda}(t)=|d_{\alpha\beta,\gamma\lambda}(t)|^2$ and  its explicit form is 
the following
\be
D_{\alpha\beta,\gamma\lambda}(t)=|\bra{\eta(\la_i)}e^{iH_{\alpha\beta}(\la_f)t}
e^{-iH_{\gamma\lambda}(\la_f)t}\ket{\eta(\la_i)}|^2.
\label{eq:df}
\ee

Now, using the density matrix $\rho_s(t)$ given in Eq.~(\ref{eq:rho_s}), one can compute the concurrence 
between the two qubits. We shall follow the Wooter's definition of concurrence given by 
\be
 C(\rho_s)=\rm max(0,\epsilon_1-\epsilon_2-\epsilon_3-\epsilon_4),
\label{eq:cnc}
\ee
where $\epsilon_i$'s are the square root of the eigenvalues in a descending  order of the 
non-Hermitian matrix $M=\rho_s\hat\rho_s$ with $\hat \rho_s$ defined as 
\be
\hat\rho_s=(\sigma^y\otimes\sigma^y)\rho_s^*(\sigma^y\otimes\sigma^y).
\label{eq:R_mat}
\ee
Therefore, one readily concludes that the concurrence between the two qubits are determined by the DF
associated with the four channels.

Let us consider a generic Hamiltonian  of a one-dimensional Ising chain 
in a site dependent transverse field $\la_n$ given by
\be
H~=~-\sum_{n}( \sigma_n^x\sigma_{n+1}^x~+~\la_n\sigma_n^z).
\label{ham1}
\ee
One can obtain the initial Hamiltonian from the above Hamiltonian (\ref{ham1}) by setting $\la_n=\la$ while for the final Hamiltonian 
$\la_n$ becomes different from $\la$ at those sites where the qubits are coupled.
For the initial homogeneous case ($\la_n=\la$),
the model in Eq.~(\ref{ham1}) has a QCP at $J=\la$ separating ferromagnetic  (FM)  and 
quantum paramagnetic (PM) phases.
Using Jordan-Wigner transformations followed by Fourier transformation for a homogeneous and periodic chain, 
the energy spectrum for the Hamiltonian in Eq.~(\ref{ham1}) is obtained as \ct{lieb61,pfeuty70}
\be
\varepsilon_q=\pm2\sqrt{(\la+\cos q)^2+\sin^2q},
\label{spectrum1}
\ee
where $q$ is the momentum which takes discrete values given by $q=2\pi m/N$ with 
$m=0\cdots N-1$ for a finite system of length $N$. 

In order to express the DF given in Eq.~(\ref{eq:df}) in the fermionic representation one has to cast the Hamiltonian 
in the above basis following Jordan-Wigner transformation. The Hamiltonian in Eq.~(\ref{ham1}) 
can be described by a quadratic form in terms of 
spinless fermions $c_i$ and $c_i^{\dagger}$ \ct{lieb61,pfeuty70}
\be
H = \sum_{i,j} \left[ c_i^\dagger A_{i,j} c_j +
\frac{1}{2} ( c_i^\dagger B_{i,j} c_j^\dagger + \mathrm{h.c.}) \right].
\label{eq:ham_fermion}
\ee
Here, $A$ is a symmetric matrix as $H$ is Hermitian 
and $B$ is an antisymmetric matrix which follows from the 
anticommutation rules of $c_i$'s. 
The elements of these matrices thus obtained are:
\begin{eqnarray}
A_{i,j}&=&-(J\de_{j,i+1}+J\de_{i,j+1})-2\la_j\de_{i,j},\cr
B_{i,j}&=&-(J\de_{j,i+1}-J\de_{i,j+1}),
\label{eq_AB}
\end{eqnarray}
where $\la_j$ is the site dependent transverse field.

The Hamiltonian (\ref{eq:ham_fermion}) can be written in the following form also
\be
H={1 \over 2} \Psi^{\dagger} {\cal{H}} \Psi,
\label{eq:ham_mat}
\ee
with $\Psi=(c^{\dagger},c)=(c_1^{\dagger}, \cdots, c_N^{\dagger},c_1, \cdots, c_N)$
and $\cal H$ is given by
\be {\cal H} = \left[ \begin{array}{cc} 
-A & -B \\
B  & A \end{array} \right]. \label{eq:ham_fermion2} \ee

The above Hamiltonian can be diagonalized in terms of
the normal mode spinless Fermi operators $\eta_k$ given by the relation~\ct{lieb61}. 
\ba
\eta_k=\sum_i(g_k(i) c_i + h_k(i) c_i^{\dagger}),
\label{eq:normal}
\ea
where $g_k(i)$ and $h_k(i)$ are real numbers; $g_k(i)$ and $h_k(i)$ are obtained from the 
real matrices $g$ and $h$, respectively.

The unitary matrix $U$ that diagonalizes the Hamiltonian given in Eq.~(\ref{eq:ham_fermion2}) can be 
constructed from $g_k(i)$ and $h_k(i)$
\be U = \left[ \begin{array}{cc} 
g & h \\
h  & g \end{array} \right]. \label{eq:dia_op} \ee

Using this Unitary operator one can also write the fermionic operator 
in terms of normal modes
\be
c_i=\sum_k(g_k(i) \eta_k + h_k(i) \eta_k^{\dagger}).
\label{eq:fermion}
\ee

In terms of the new operators $\eta_k$, the Hamiltonian in Eq. \eqref{eq:ham_fermion2} takes 
the diagonal form,
\be
H = \sum_k \Lambda_k \left( \eta_k^\dagger \eta_k - \frac{1}{2} \right),
\ee
with $\Lambda_k$ being the energy of different fermionic modes with index $k$. 

In order to study the time evolution of concurrence $C$, we have  to first compute the time-dependent DF that constitutes
the reduced density matrix of two spins given in Eq.~(\ref{eq:rho_s}). 
One can the use the covariance matrix formalism to determine time evolution of DF 
associated with the different DCs  governed by two different Hamiltonians \ct{wendenbaum14,keyl10}.
These different Hamiltonians $H_{\alpha\beta}$ shown in Eq.~(\ref{eq:channel}) are having different set of local transverse fields. 
This formalism allows us to write
the DF in the following way 
\be
D_{\alpha\beta,\gamma\lambda}(t)=|{\rm det}(I-R_{\alpha\beta}(t)-R_{\gamma\lambda}(t))|^{1/2}.
\label{eq:echo1}
\ee
Here, $I$ is an identity matrix and $R_{\alpha\beta}(t)$'s are the time evolved covariant matrices given by 
\be
R_{\alpha\beta}(t)=e^{-iH_{\alpha\beta}t}R(0)e^{iH_{\alpha\beta}t}
\label{eq:echo2}
\ee
with $R(0)=\bra{\eta(\la_i,t=0)}\Psi\Psi^{\dagger}\ket{\eta(\la_i,t=0)}$, and its matrix form is given by 

\be
 R(0) = \left[ \begin{array}{cc} 
\langle c^{\dagger}c\rangle & \langle c^{\dagger}c^{\dagger}\rangle \\
\langle cc\rangle  & \langle c c^{\dagger}\rangle \end{array} \right]
=\left[ \begin{array}{cc} 
h^T h & h^T g \\
g^T h  & g^T g \end{array} \right],
\label{eq:echo3}
\ee
where $T$ denotes the transpose of a matrix.
This $2N \times 2N$ initial 
covariant matrix is composed of four blocks and all of these blocks are having the same dimension of $N\times N$.

%

\section{Equilibrium study}
\label{result1}

\begin{figure}[ht]
\begin{center}
\includegraphics[height=5.0cm,width=4.1cm]{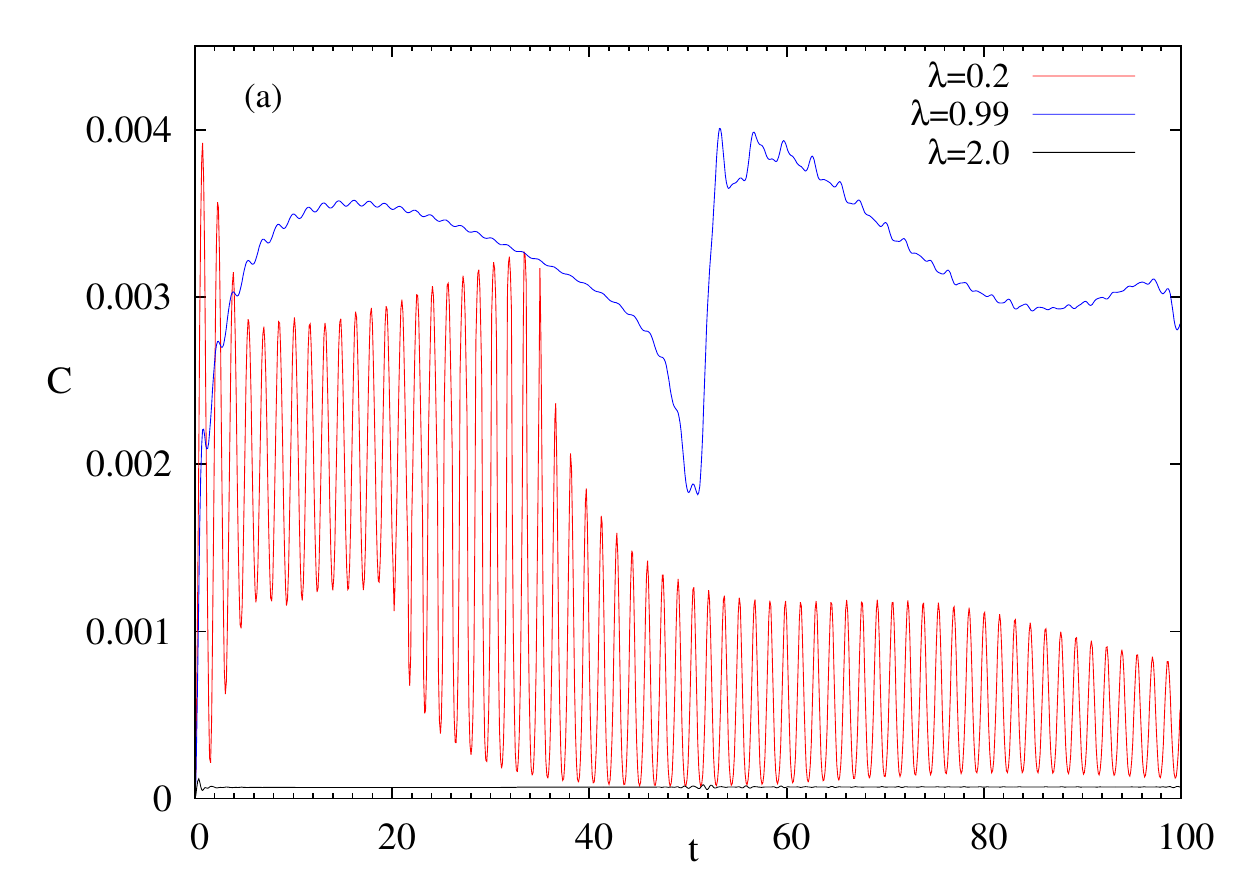}
\includegraphics[height=5.0cm,width=4.1cm]{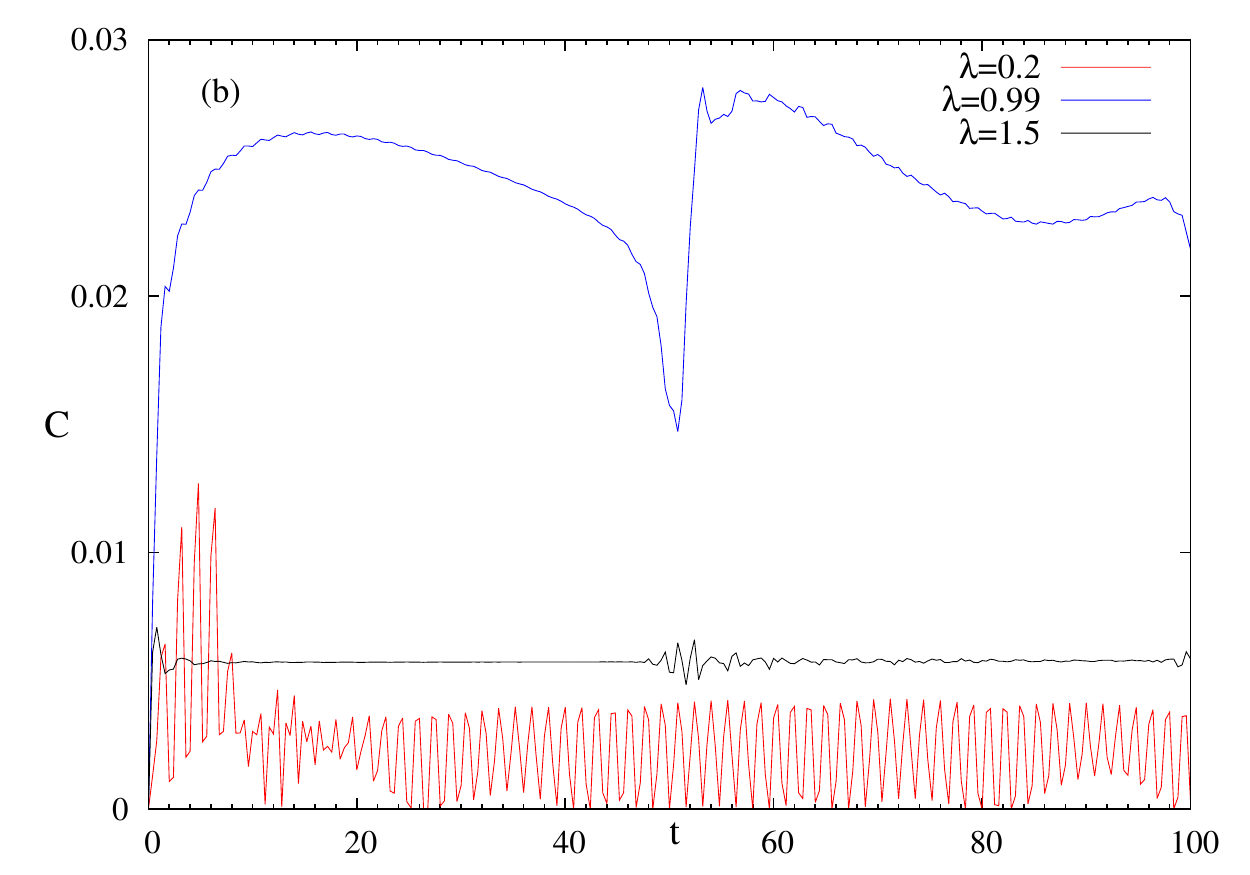}
\end{center}
\caption{(Color online) Plot (a) shows the equilibrium behavior of concurrence $C$ as a function of time  with both  the qubits 
 connected at the same site i.e., $d=0$, for  different  phases (FM and PM phase) including the QCP of the environmental chain; 
Plot (b) shows the variation of $C$ as a function of time when $d=1$. In both the cases,  the  generation of entanglement is of small magnitude.
We consider $N=100$ and $\delta=0.1$.} \label{fig1} \end{figure} 

\begin{figure}[ht]
\begin{center}
\includegraphics[height=5.0cm,width=7.1cm]{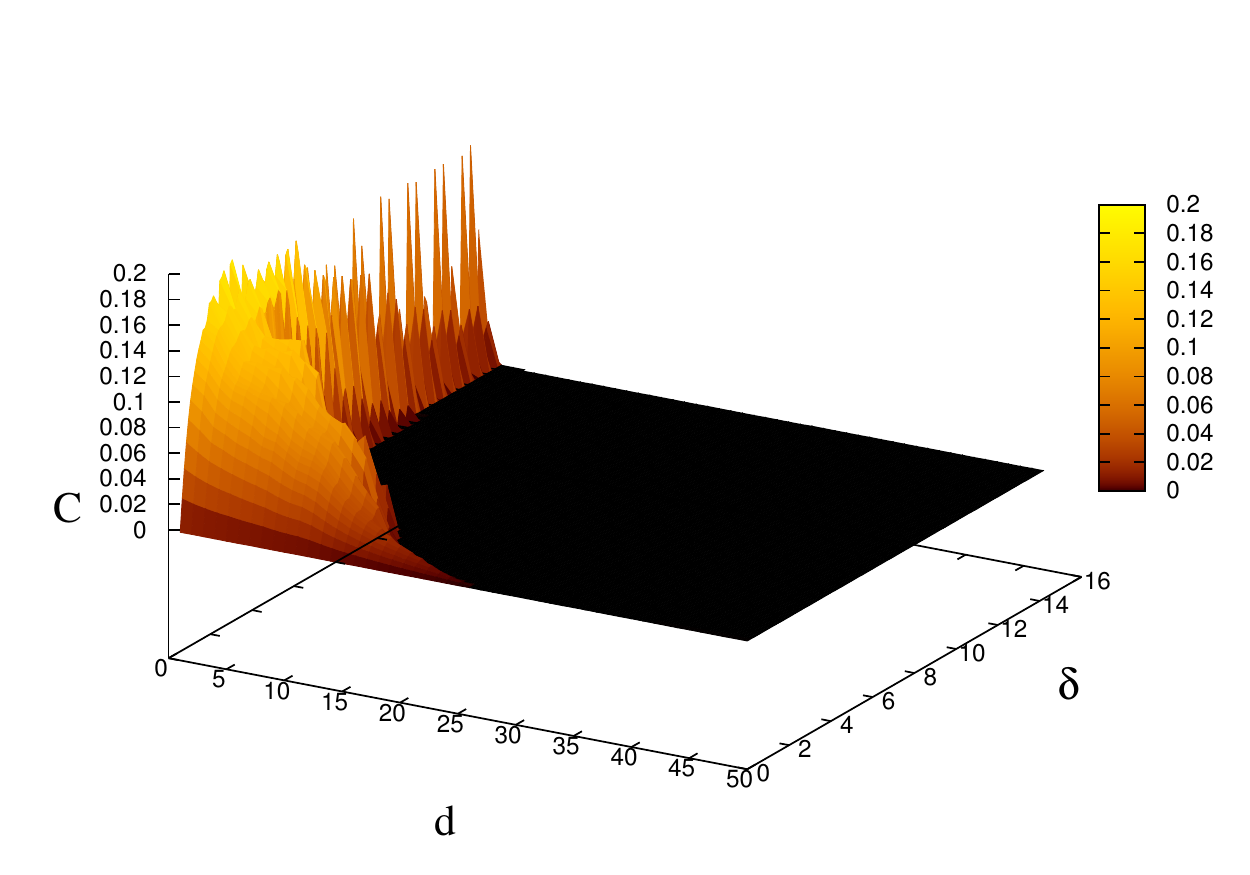}
\end{center}
\caption{(Color online) Plot shows the variation of $C$ as a function of distance $d$ and coupling strength $\delta$ when
$\lambda=0.99$ (QCP), $N=100$, 
and $t=7.3$. In the weak coupling limit the concurrence is non-zero for relatively larger  separation $d$. } \label{fig2} \end{figure}

In this section, we shall illustrate the equilibrium behavior of concurrence given in Eq.~\eqref{eq:cnc} as a function of time 
when the environment evolves along different DCs originated from the coupling to the qubit.
In this case $\la_i=\la_f=\la$, for all the sites except the sites where the qubits are locally connected. Figure~(\ref{fig1}a) depicts the behavior of 
concurrence for $d=0$ while Fig.~(\ref{fig1}b)) shows it for $d=1$.
 When the parameter value is chosen to be close to the
quantum critical value ($\la=0.99$), the concurrence initially grows as a function of time showing a prominent dip
at $t=N/v=N/2$; this is because  of the constructive interference of quasi-particles
 generated due to the local connection of the qubit to the spin chain having group velocity $v=2$ at the QCP.
Thus, the finite size effect is manifested in this dip of the LE which is prominent for
$\la=1$.  In the PM phase also, concurrence shows rapid fluctuations of small amplitude  at around  $t=N/2$;  otherwise, 
it shows a time independent behavior. 
What is noteworthy that even if the qubits are initially unentangled, there is a generation concurrence  only due to the local coupling  during
the temporal evolution of the composite system.

Additionally, maximum concurrence is generated when $\la$ is close to the critical
value (when there is a diverging length scale)  rather than  in the PM and FM phase.
  That the value of the concurrence attains
 maximum  for $\la \simeq 1$ is independent of the distance $d$ between the qubits and hence is a universal
 observation. We however note that the the magnitude of the maximum value of the concurrence thus generated is very small in the
 equilibrium case in contrast to the non-equilibrium case to be discussed in the next section.

 Furthermore,
 a non-zero concurrence survives when $d$ increases within the small $\delta$ limit; on the other hand, in the large
 $\de$ limit, concurrence is smaller compared to the small $\de$ case (see Fig.~(\ref{fig2})). 
A large value of $\delta$ makes the value of the local transverse field  off-critical 
so that  the correlation become short ranged and consequently the entanglement between two distant qubits is vanishingly small while
for small $\de$ the value of the local transverse field stays critical and hence a long-range correlation exists in the environment.

\section{Non-equilibrium study}
\label{result2}
\begin{figure}[ht]
\begin{center}
\includegraphics[height=4.5cm,width=4.0cm]{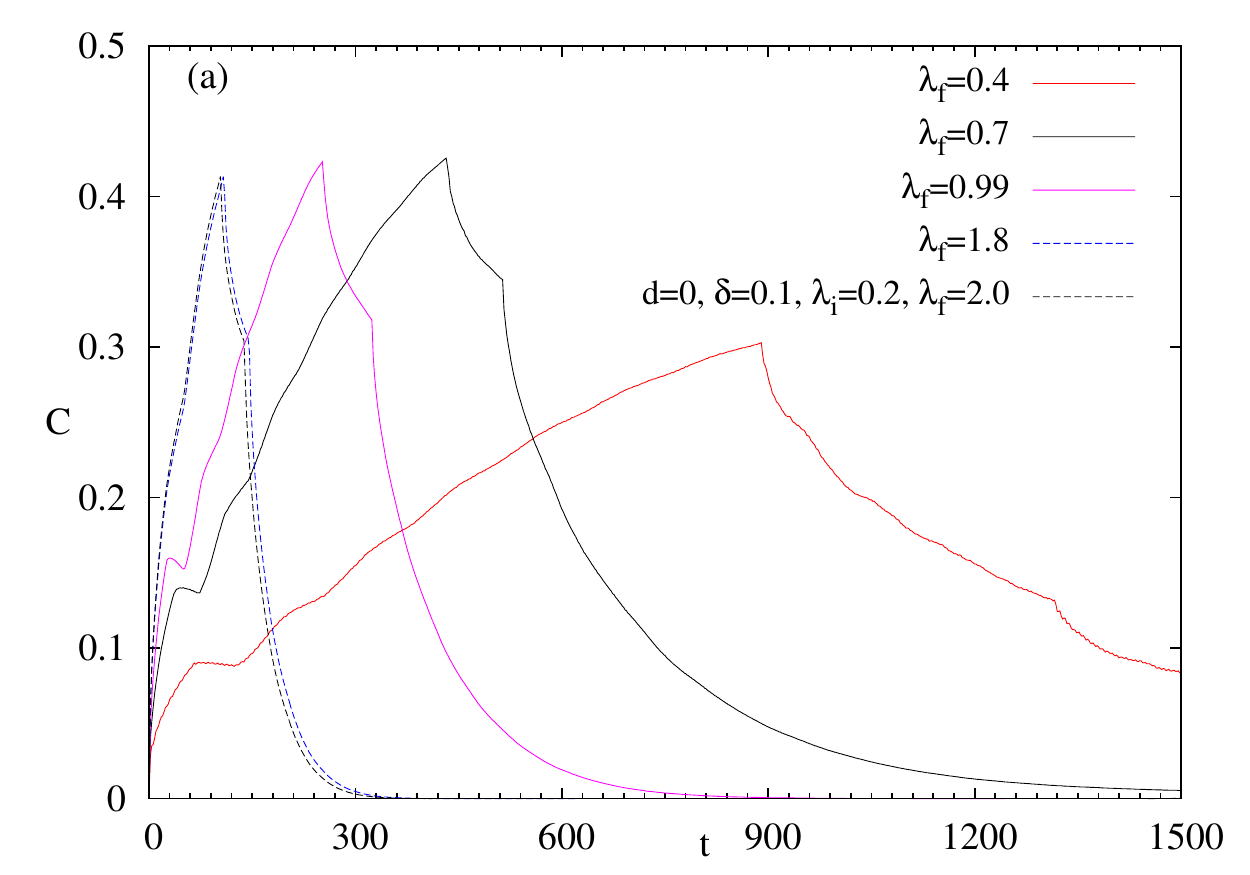}
 \includegraphics[height=4.5cm,width=4.0cm]{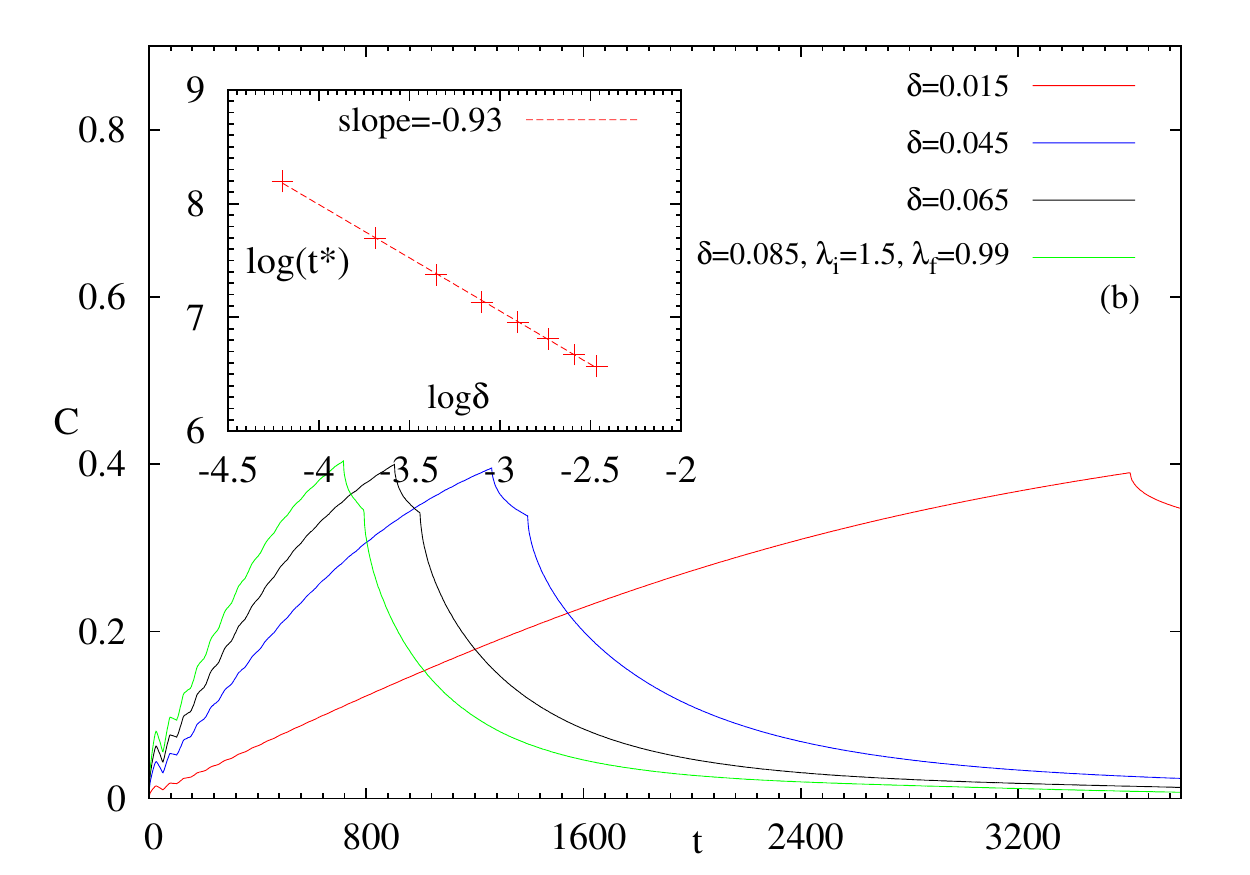}
\end{center}
\caption{(Color online) Plot (a) shows the variation of $C$ as a function of time followed by a sudden quench of the transverse field of the 
environmental Ising chain choosing
the environment initially in the FM phase. The concurrence initially increases and eventually decays
to zero showing a primary peak at $t=t^*$ and a secondary peak at $t=t_2>t^*$; $t^*$ and $t_2$ both decrease with the quench amplitude $|\la_i-\la_f|$. Plot (b) shows the variation of $C$ as
a function of time when the spin chain is quenched from the PM phase with $\la_i=1.5$ to  QCP at $\la_f=0.99$ for different values of $\delta$.
 Inset shows the variation of $t^*$ with  $\delta$ within the weak coupling limit, $t^*\propto \delta^{-1}$.
 For both the above cases, we consider $d=0$.
} \label{fig3} \end{figure}

We shall now extend the previous studies to the situation in which the environmental spin chain undergoes
a global sudden quenching,  i.e., the transverse field $\la$ is suddenly changed from an initial
value $\la_i$ to a final value $\la_f$. In this non-equilibrium situation, two external qubits become more strongly entangled as 
compared to the earlier equilibrium situation.  Results presented in Fig. (\ref{fig3}a) suggest: (i)  the concurrence generation is maximum when
the spin chain is quenched to the QCP starting from the FM phase. (ii)  Quenching 
within the same phase yields entanglement of smaller magnitude  between the qubits as compared to the quenching between two 
different phases. Furthermore, concurrence remains non-zero 
for longer time if the quenching is performed within the same phase.
Finally, there is a prominent peak in $C$
appearing at time $t=t^*$ that becomes smaller for higher quench amplitude. Additionally, there exists a secondary
peak at $t=t_2$ after the primary peak at $t^*$. 
On the other hand, we show in 
Fig.~(\ref{fig3}b) that $t^*$ decreases with $\de$, in fact, is inversely proportional to $\delta$ as shown in the inset.
We note that the above features of $C$ is qualitatively identical  for all types of  quenching protocols.

\begin{figure}[ht]
\begin{center}
\includegraphics[height=4.5cm,width=7.0cm]{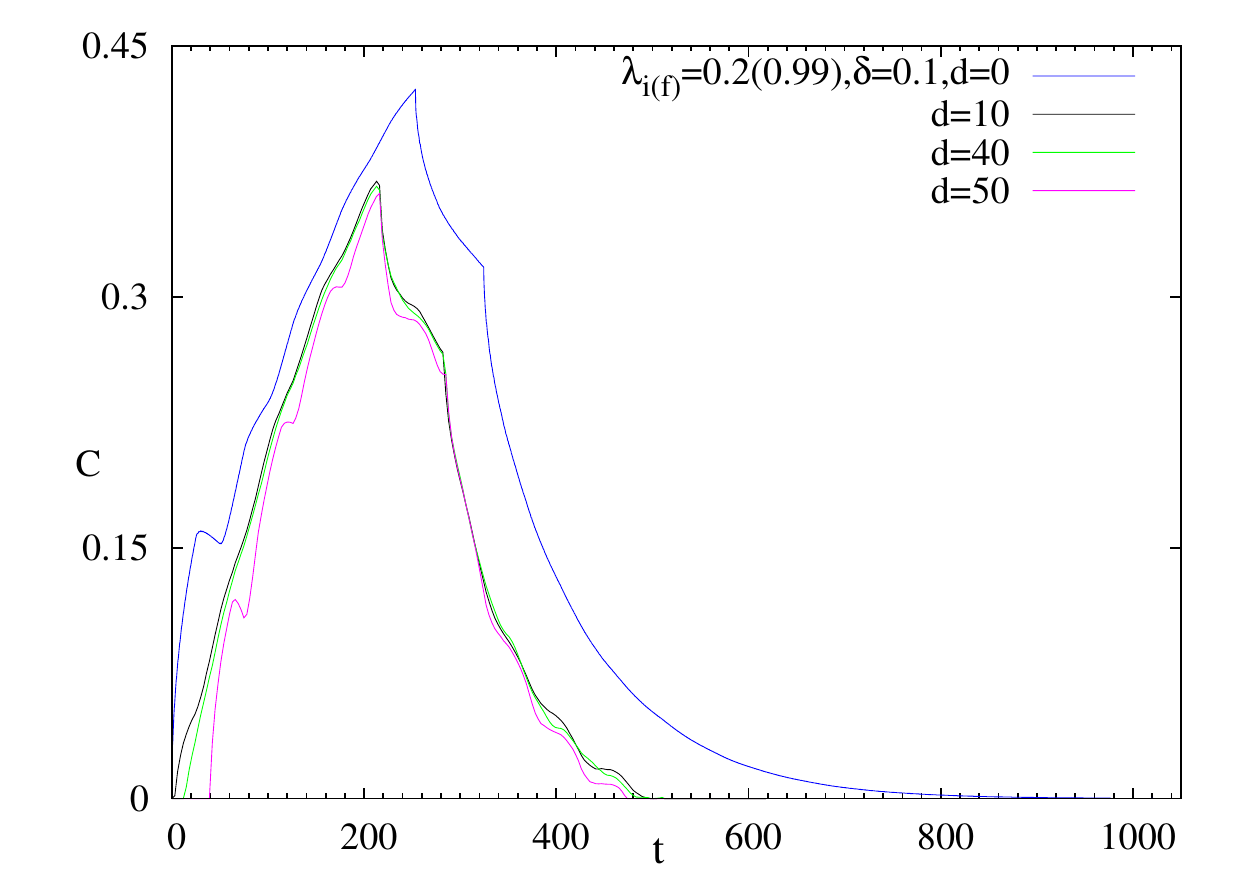}
\end{center}
\caption{(Color online) Figure shows  the variation of $C$ as a function of time with distance $d$ as the parameter 
when the environmental chain is quenched from the FM phase to QCP. The concurrence becomes maximum for 
$d=0$; there exists a  threshold time $t_{\rm TH}$ above which $C$ attains a finite value for 
$d \neq 0$.} 
\label{fig4} \end{figure}

Figure (\ref{fig4}) shows that for the critical quenching starting from the FM phase, $C$ becomes maximum for 
$d=0$ while for other cases with $d\neq 0$, it attains a finite value only after a threshold time $t_{\rm TH}$; this threshold time 
increases with the increasing $d$.
 One can note that $t^*$ attains a higher value for $d=0$ as compared to the case $d\neq 0$;  in the latter case,
 $t^*$ almost remains constant. Additionally, we observe that $C$ stays at non-zero vale for longer time for $d=0$ as compared to $d\neq 0$ case.

\section{Interpretation using channel analysis}

\begin{figure}[ht]
\begin{center}
\includegraphics[height=6.5cm]{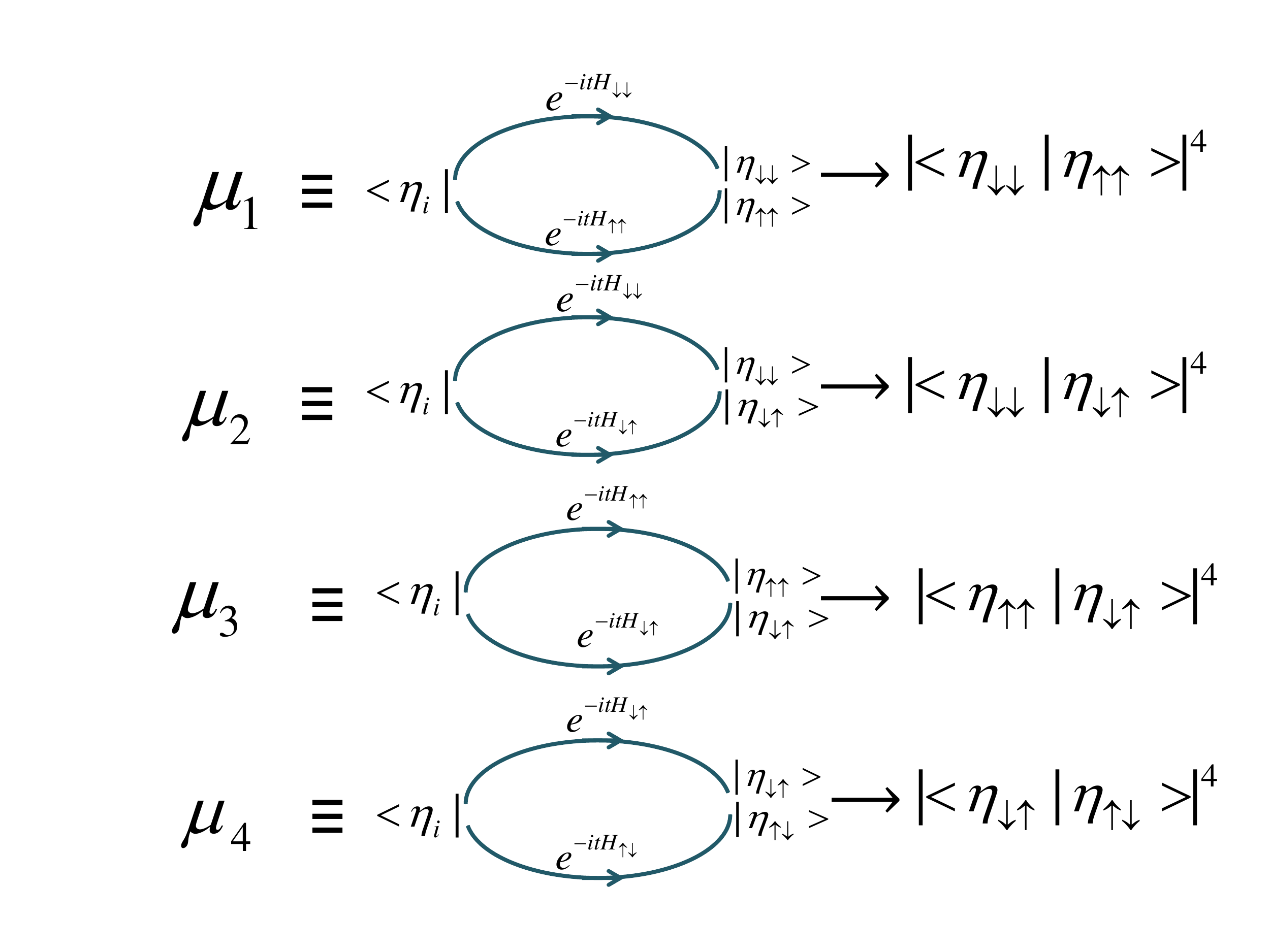}
\end{center}
\caption{(Color online) Schematic diagram explicitly shows the formation of four channels $\mu_1$, $\mu_2$, $\mu_3$ and $\mu_4$ with their 
underlying Hamiltonians $H_{\alpha \beta}$; $\mu_1$ is governed by $H_{\uparrow\uparrow}$ and $H_{\downarrow\downarrow}$,
while $\mu_2$ by $H_{\downarrow\downarrow}$ and $H_{\downarrow\uparrow}$,  
$\mu_3$ by $H_{\uparrow\uparrow}$ and $H_{\downarrow\uparrow}$ and
$\mu_4$  by $H_{\downarrow\uparrow}$ and $H_{\uparrow\downarrow}$.
} \label{fig11} \end{figure}

In this section, we shall analyze the results presented in previous sections using the DF (or the LEs) associated with the different
 DCs  which in turn lead to the generation of entanglement between the qubits which are initially
unentangled.   For this purpose,
let us fix our notation first:  $L(\alpha \beta,\gamma \lambda)=|{\rm det}(I-R_{\alpha \beta}-R_{\gamma \la})|=|d_{\alpha\beta,\delta\gamma}|^4$;
$L(\downarrow \downarrow, \uparrow\uparrow)=\mu_1$; periodic boundary condition ensures that
$L(\downarrow \downarrow,\uparrow \downarrow)=L(\downarrow \downarrow,\downarrow\uparrow )=\mu_2$,
$L(\uparrow\uparrow,\uparrow \downarrow)=L(\uparrow\uparrow, \downarrow \uparrow)=\mu_3$ (hence, this is valid for all $d$
as well as  for equilibrium and non-equilibrium cases) and 
$L(\downarrow\uparrow,\uparrow \downarrow)=\mu_4$.  We therefore have to deal with these four DCs to analyze the temporal behavior of $C$. We shall refer $\mu_1,~\mu_2,~\mu_3$ and $\mu_4$ as DCs  in 
the subsequent discussion and figures. The schematic diagram as shown in Fig.~(\ref{fig11}) depicts different Hamiltonians
$H_{\uparrow\uparrow}$, $H_{\uparrow\downarrow}$, $H_{\downarrow\uparrow}$ and $H_{\downarrow\downarrow}$ that govern the time evolution of 
four DCs.

\begin{figure}[ht]
\begin{center}
\includegraphics[height=6.50cm,width=4.20cm]{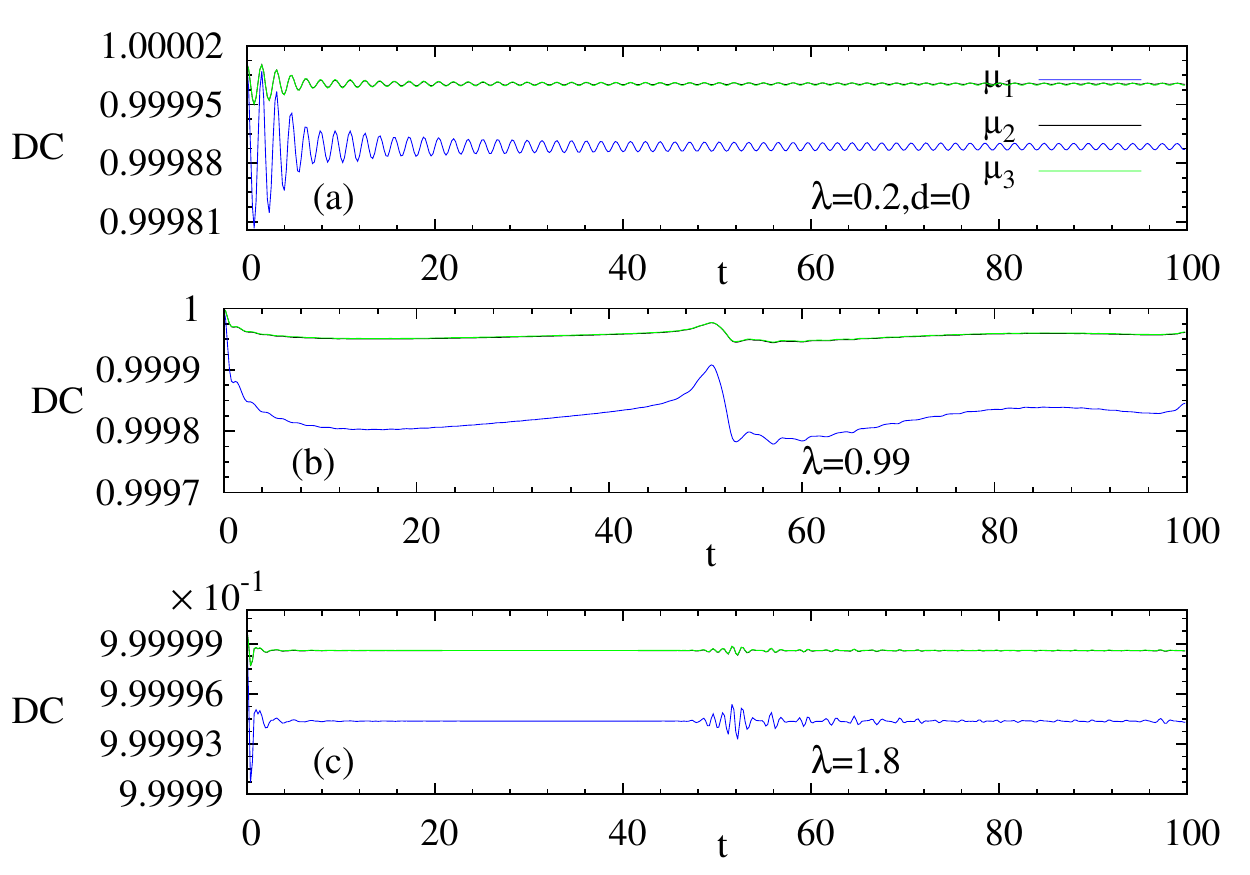}
\includegraphics[height=6.50cm,width=4.20cm]{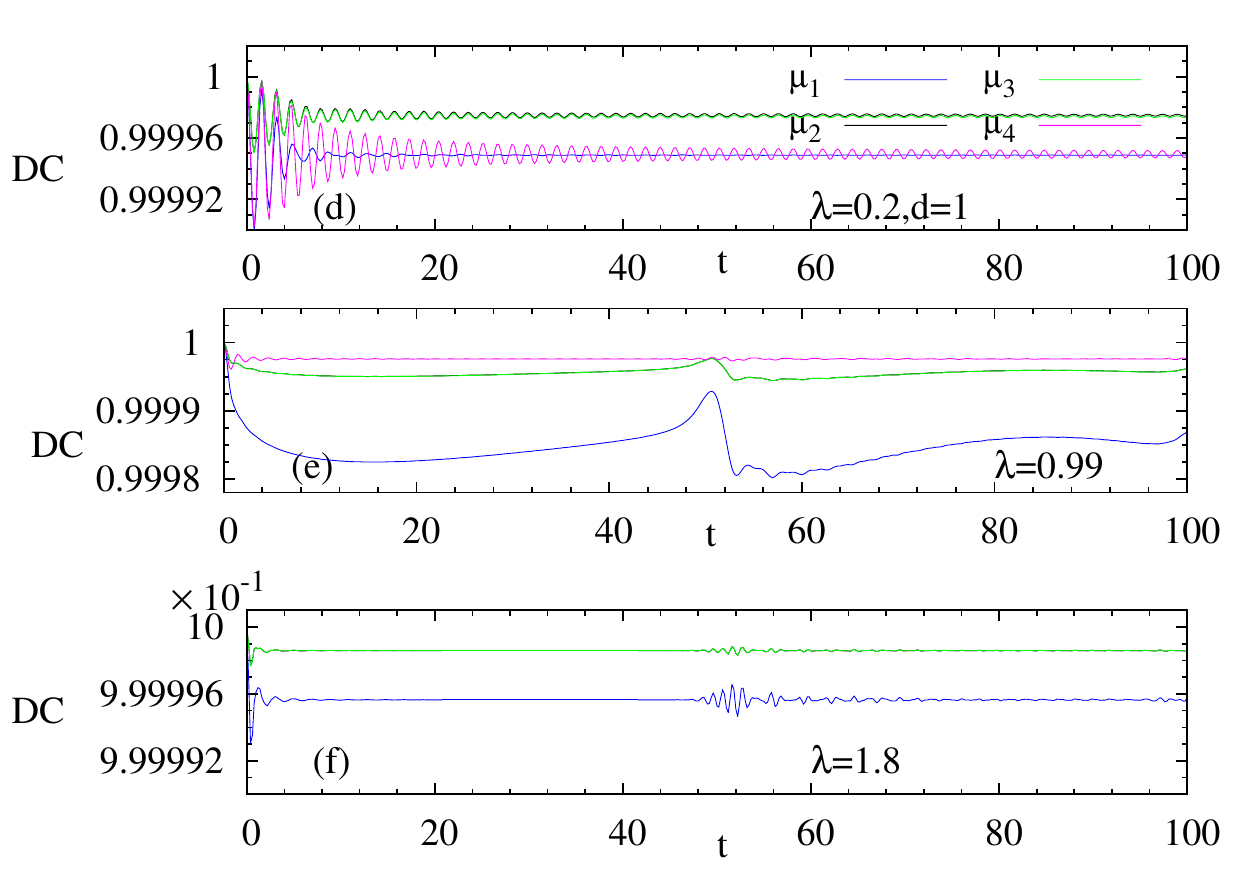}
\end{center}
\caption{(Color online) Equilibrium echo for different DCs $\mu_1$, $\mu_2$ and  $\mu_3$ as a  function of time for $d=0$ are plotted
when the spin chain is in the FM Phase (a),  at the QCP (b)  in the PM phase (c).
We observe that $\mu_3$ superimposes on $\mu_2$ leading to   
 two non-trivial independent channels as explained in the text. 
 The above DCs along with $\mu_4$ are studied for the situation $d=1$ in the following plots.
 There is an additional independent channel $\mu_4$ which is no longer trivially unity like earlier $d=0$ case. 
Plot (d) is for the FM phase; 
 the plot (e) and the
plot (f) represent the situations when spin chain is quantum critical and in the PM phase, respectively.
For both the cases we consider, $N=100$, $\delta=0.01$.
} \label{fig5} \end{figure}

Let us first investigate the behavior of individual channel in equilibrium scenario.
When two qubits are coupled at the same 
site of the environmental chain i.e., $d=0$, only three of the four channels mentioned above are  independent with  corresponding DFs   $\sqrt{\mu_1}$, $\sqrt{\mu_2}$ and 
$\sqrt{\mu_3}$. Figure (\ref{fig5}a,b,c) represent the temporal behavior of different DCs for $d=0$ when the environment is in the FM phase, at QCP and in the PM phase, respectively. The other channel $\mu_4$ does not have a dynamics and hence  remains trivially  unity for all time due to the  fact that the
initial state evolves with two identical Hamiltonians both with an additional transverse field $\delta$ at one site. Furthermore, in 
the weak coupling limit $\delta \to 0$,
$\mu_2 \simeq \mu_3$; therefore, 
one can approximately work with  two independent DCs, $\mu_1$ and $\mu_2$. 
These observations lead us to the  conclusion that
for $d=0$ and within the weak coupling limit, the number of independent 
channels effectively depends the corresponding difference of in the local fields of two Hamiltonians those dictate the time
evolution of the initial state; the differences in this case are $\delta$ for $\mu_2$ and $2 \delta$ for $\mu_1$. The other notable point in Fig.(\ref{fig5}a,b,c) is
$\mu_2$, calculated in different phases and at QCP, always remains at a higher value than that of the $\mu_1$.
This can be attributed to the fact that 
$\mu_1$ exhibits a sharper short time fall than that of $\mu_2$. 
This is a manifestation of the  difference in
local transverse field; in  $\mu_1$ it  is of the order $2\delta$ while $O(\delta)$ in $\mu_2$. 

Let us now proceed to the case for $d\neq 0$ and $\delta\to 0$, where indeed we have three independent channels.
Figure (\ref{fig5}d,e,f) represent the temporal behavior of different DCs for $d=1$ when the environment is in the FM phase, at the QCP and in the PM phase, respectively.
Interestingly,  $\mu_4$ in this case is not trivially unity as in the previous case of $d=0$; additionally,  
 $\mu_2$ deviates from $\mu_3$ as the coupling strength  $\delta$ increases  resulting in
four independent channels for higher values of $\delta$. One can see that in the FM phase $\mu_2$ is always higher than $\mu_1$ and $\mu_4$ as shown 
in  Fig.(\ref{fig5}d,e,f). On the other hand, in the PM phase or at the QCP $\mu_4>\mu_2>\mu_1$. One can hence conclude that for $d\ne 0$ the 
channel $\mu_4$ is maximally affected in the FM phase as compared to the QCP and PM phase. In this FM phase, $\mu_1$ and $\mu_4$ 
almost overlap with each other;
 $\mu_1$ in the above phase shows prominent oscillations as one increases $d$. In all the cases discussed above, $\mu_1$
is maximally deviated from unity during its temporal evolution.

One can see from all cases shown the  Fig.~(\ref{fig5}) that the DC $\mu_2$ almost coincides
with $\mu_3$. This can be attributed to fact that both of them are governed by the Hamiltonians which are deviated from each other in an 
identical way i.e., the local transverse field of the two underlying Hamiltonians are deviated by $\delta$ from each other.
This is valid irrespective of the distance between the two qubits $d$.

To compare the dephasing rate of a single qubit \ct{yu03} with the temporal decay of the concurrence which is generated following a non-equilibrium evolution,
we invoke upon the reduced density matrix that is obtained by tracing over one of the qubits from the density matrix of two qubits (\ref{eq:rho_s}):
\be \rho_A(t) =  \rho_B(t)= {1\over 2}\left[ \begin{array}{cc} 
1 & { \sqrt{\mu_3} + \sqrt{\mu_2} \over 2} \\
{ \sqrt{\mu_3} + \sqrt{\mu_2} \over 2} & 1 
\end{array} \right]. \label{eq:rho_sa} \ee
Here, $S_{A(B)}$ denotes the dephasing factor (DP), incorporated in  the off-diagonal terms of
$\rho_{A(B)}$, of the qubit $A(B)$. {DP quantifies the loss of coherence  of a single qubit which was initially 
prepared in a pure state with another qubit; the decay time  is  determined by the dephasing rate.} Remarkably, $S_{\rm A}=S_{\rm B}=(\sqrt{\mu_3}+\sqrt{\mu_2})/4$ and is completely independent of  $\mu_1$ and $\mu_4$; this leads to an interesting consequence
as we shall elaborate below.

\begin{figure}[ht]
\begin{center}
\includegraphics[height=5.50cm,width=7.1cm]{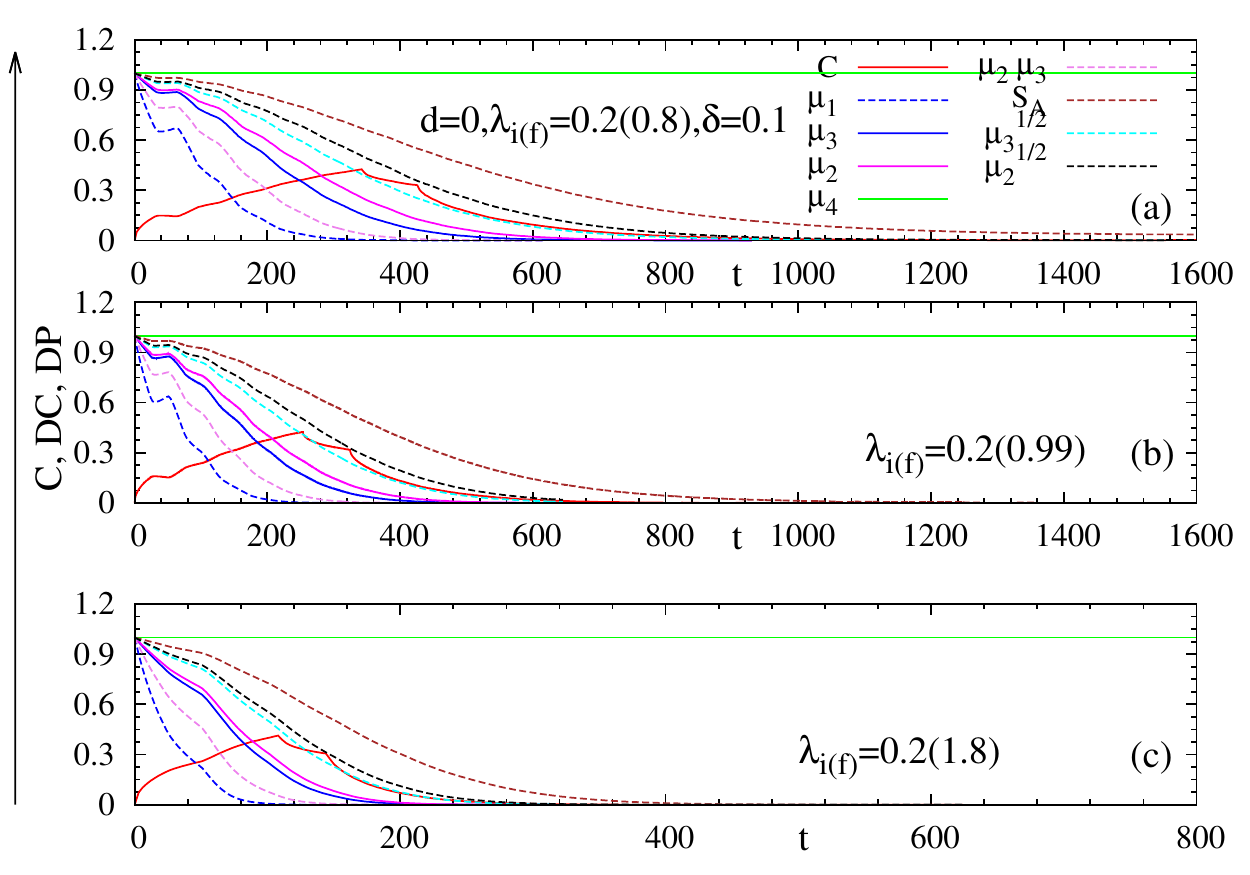}
\end{center}
\caption{(Color online) Non-equilibrium echos for different DCs, DP and $C$  are plotted as a function of time with three different
situations, quenching to the FM phase with $\la_f=0.8$ (a), quenching to the QCP with $\la_f=0.99$ (b) and quenching to the PM phase where $\la_f=1.8$ (c).
Here, $N=100$, $\delta=0.1$, $d=0$ and $\la_{i}=0.2$.
$\mu_1$ displays the sharpest decay to zero among all the four channels. $\mu_4$ exhibits time independent behavior and $\mu_3$ vanishes more rapidly than $\mu_2$.
} \label{fig6} \end{figure}

Now, we focus on the non-equilibrium evolution of the channel choosing  $d=0$, first. Here, one has three  independent channels unless $\delta \to 0$
when $\mu_2$ coincides with $\mu_3$;   $\mu_4$ becomes trivially identity as in the equilibrium case. Figure.~(\ref{fig6}a,b,c) show 
non-equilibrium temporal evolution of $C$, DC and DP with $d=0$ for the FM, the critical and the PM quenching while the initial state of 
the environment is in the FM phase, respectively. One
can see that
the primary peak of the 
concurrence occurs when $\mu_1 \to 0$ and the secondary peak when 
$\mu_2 \mu_3 \to 0$. On the other hand, concurrence becomes vanishingly small when $\mu_3^{1/2}\to 0$.
Additionally, one can see that in the early time region $\mu_2$ almost overlaps
with $\mu_3$ but in the course of the evolution, these two channels starts to behave differently leading to a visible deviation from each other.
Recalling $S_{A(B)}$ we find  that there exists a finite coherence even long after the qubits become
unentangled from each other; this implies that the dephasing rate is slower than the rate in which 
the qubits lose the entanglement\ct{yu03}. 
This observation is qualitatively explained as follows: the DCs  $\mu_2$ and $\mu_3$ appear in the DP $S_A$ and $S_B$  in an additive manner; on the other hand, concurrence 
depends on the $\mu$'s in a complicated way. We therefore observe a long dephasing time $T_{\rm D}$ (dictating the decay of $S_A$ and $S_B$)
as compared to unentanglement time $t_{\rm UE}$ above which the concurrence vanishes between the two qubits.
Additionally,  $T_{\rm D} \gg T_{\rm UE}$ for FM quenching 
where as  $T_{\rm D}$ is  comparable (of the same order) to  $T_{\rm UE}$ for the PM quenching case and the quantum critical case.
All the above observations are independent of the quenching path i.e., sudden 
quenching within the same phase or to QCP or to a different phase. 

\begin{figure}[ht]
\begin{center}
\includegraphics[height=5.50cm,width=7.1cm]{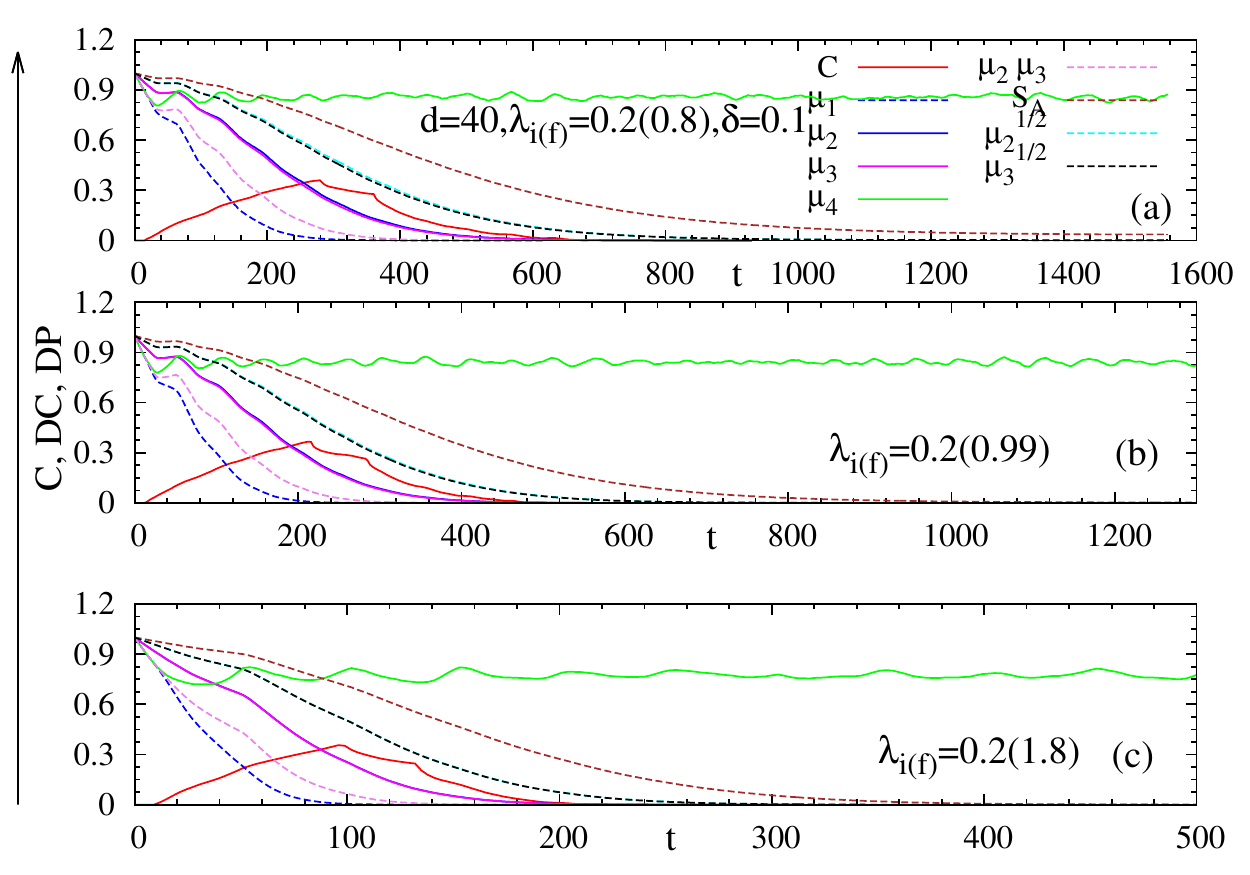}
\end{center}
\caption{(Color online) DC, DP and $C$ for $d=40$ are plotted as a function of time with three different situations, 
quenching to the FM phase with $\la_f=0.8$ (a), quenching to QCP with $\la_f=0.99$ (b), quenching to the PM phase where $\la_f=1.8$ (c).
Here, $N=100$, $\delta=0.1$, and $\la_{i}=0.2$.
There exists a threshold time $t_{\rm TH}$ above which $C$ attains a non-zero value. 
The dynamical behavior of $\mu_2$ coincides with that of the $\mu_3$ through out the temporal evolution.
} \label{fig7} \end{figure}

In parallel, Fig.~(\ref{fig7}a,b,c) represent the time evolution of $C$, DC and DP with $d=40$ 
for the FM, critical and the PM quenching while the initial state in the FM phase, respectively.
For this $d\neq 0$ case,
the $\mu_4$ exhibits improper oscillations with time. $\mu_4$ remains the minimally affected channel in the non-equilibrium situations like
the equilibrium situations as it always lies close to unity.
But the main difference with $d=0$ case is that $\mu_3$ and $\mu_2$ almost always coincide
with each other even when $\delta$ is not vanishingly small. Therefore, here we have three independent channels $\mu_1$, $\mu_2$ 
and $\mu_4$.  It shows that
the primary peak of $C$ occurs when $\mu_1 \to 0$ and the secondary peak is obtained when 
$\mu_2 \mu_3 \simeq \mu_2^2 \to 0$. The concurrence becomes vanishingly small when $\mu_3\to 0$.
In all the phases $\mu_2^{1/2}$ and $\mu_3^{1/2}$ remain finite even after concurrence vanishes.
The other notable feature of this finite 
$d$ case is that the $T_{UE} \simeq T_D/2$ for critical and PM quenching. 
$T_D\gg T_{UE}$ for FM quenching case which has also been observed for $d=0$ 
situation. 
 {One remarkable observation for $d\neq 0$ is that up to a threshold time $t_{\rm TH}$, concurrence 
 remains zero and different DCs  overlap with each other; for $t<t_{\rm TH}$, we see that $\mu_4\simeq \mu_1$ and $\mu_3^2\simeq\mu_1$ and 
 after this threshold time the different
 DCs move away from each other except for the channels $\mu_2$ and $\mu_3$. $C$ can only take a positive value after $t_{\rm TH}$ as shown
 in Fig.~(\ref{fig4}).}

Comparing the results presented in Figs. (\ref{fig6}) and  Figs. (\ref{fig7}), we note  
that for $d=0$, $H_{\downarrow\uparrow}$ determining the evolution of $\mu_2$ has a local transverse field modified by only $\delta$ with respect to the final unperturbed 
Hamiltonian $H_{\downarrow \downarrow}$ in which there is effect of coupling; on the other hand, $H_{\uparrow\uparrow}$ in $\mu_3$ has a local field modified by $2\delta$. Hence, $\mu_2$ and $\mu_3$ behave differently with time. 
One can infer that the dynamical evolution of $\mu_2$ matches with that of the $\mu_3$ only in the $\delta \to 0$ limit as we have already mentioned earlier.
Now, for $d\neq 0$, the underlying Hamiltonians $H_{\downarrow\uparrow}$, $H_{\uparrow\uparrow}$ governing the dynamics of $\mu_2$ and $\mu_3$ are 
 similar in the sense that both of them are having the 
identically modified  transverse fields by just an amount $\delta$ at two different sites over the Hamiltonian $H_{\downarrow \downarrow}$.
This explains  the observation that the temporal evolution $\mu_2$ and $\mu_3$,
are identical and they fall on top of each other with time.

One can write the composite density matrix of two qubits  in a generic form, valid for equilibrium as well as
non-equilibrium situations, is given by
\be \rho_s(t) = {1\over 4}\left[ \begin{array}{cccc} 
1 & \sqrt{\mu_3} &\sqrt{\mu_3} & \sqrt{\mu_1} \\
\sqrt{\mu_3} & 1 & \sqrt{\mu_4} & \sqrt{\mu_2} \\
\sqrt{\mu_3} & \sqrt{\mu_4} & 1 & \sqrt{\mu_2} \\
\sqrt{\mu_1} & \sqrt{\mu_2} & \sqrt{\mu_2} & 1
\end{array} \right]. \label{eq:rho_s1} \ee
The above density matrix can be reduced to a simplified form when $d\ne0$ with  $\mu_2=\mu_3$. 
 On the other hand, when $d=0$ and $\delta\to 0$ one has $\mu_4=1$ and $\mu_2 \to \mu_3$, respectively.

The four eigenvalues obtained from $\rho_s \hat \rho_s$ with $\mu_2=\mu_3$  are given by
\ba
\epsilon_1&=&(-1+ \sqrt{\mu_4})^2, \epsilon_2=(-1+\sqrt{\mu_1})^2, \non \\
\epsilon_{3,4}&=&{1 \over 2}\biggl( 2-8\mu_3+2\sqrt{\mu_4}+ \mu_4+2\sqrt{\mu_1}+\mu_1 \mp\non \\
&&\sqrt{(\sqrt{\mu_4}-\sqrt{\mu_1})^2(-16\mu_3+(2+\sqrt{\mu_4}+\sqrt{\mu_1})^2)} \biggr)
\label{eq:eigen2}
\ea

\begin{figure}[ht]
\begin{center}
\includegraphics[height=2.50cm,width=7.1cm]{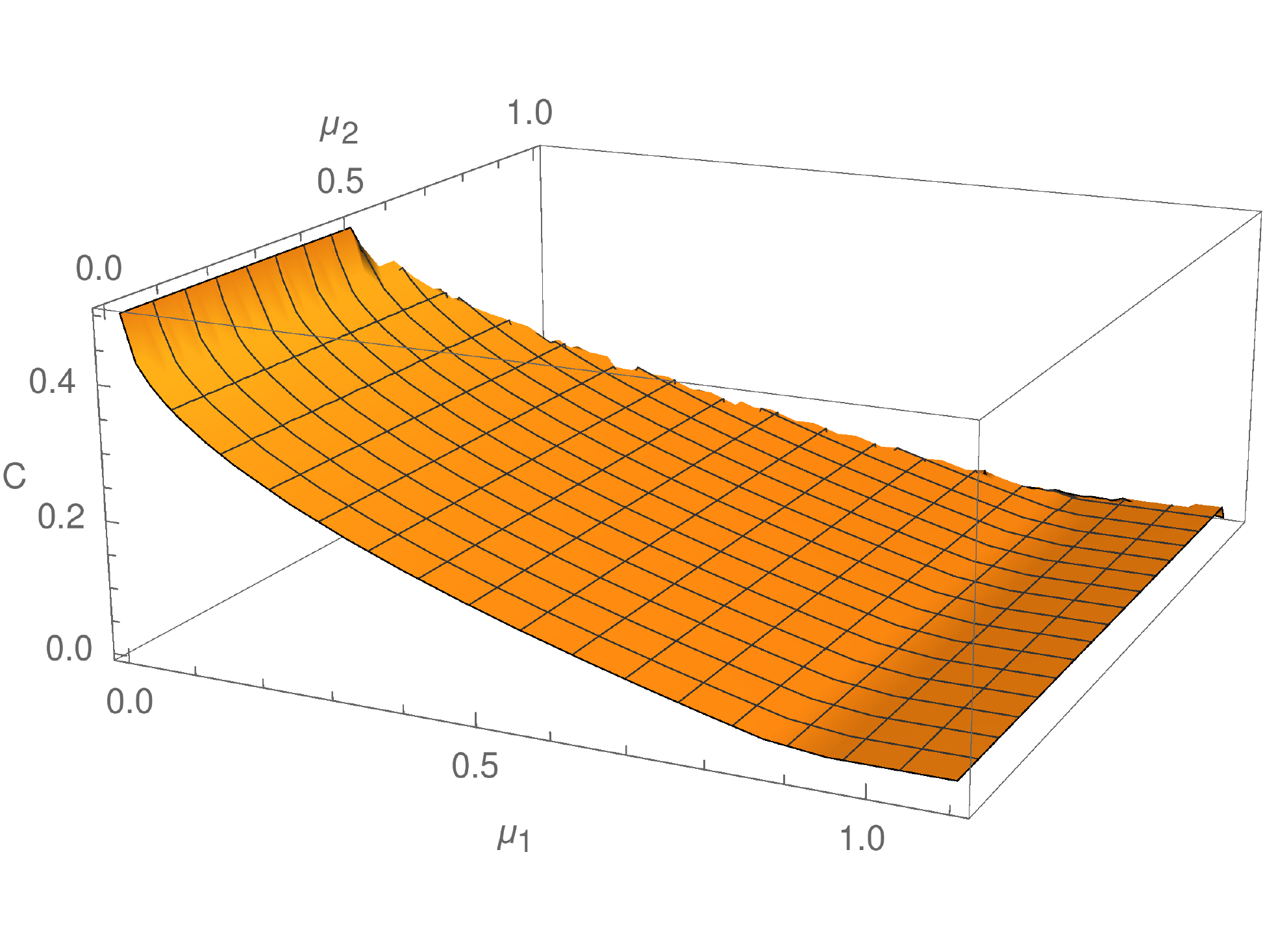}
\includegraphics[height=2.50cm,width=7.1cm]{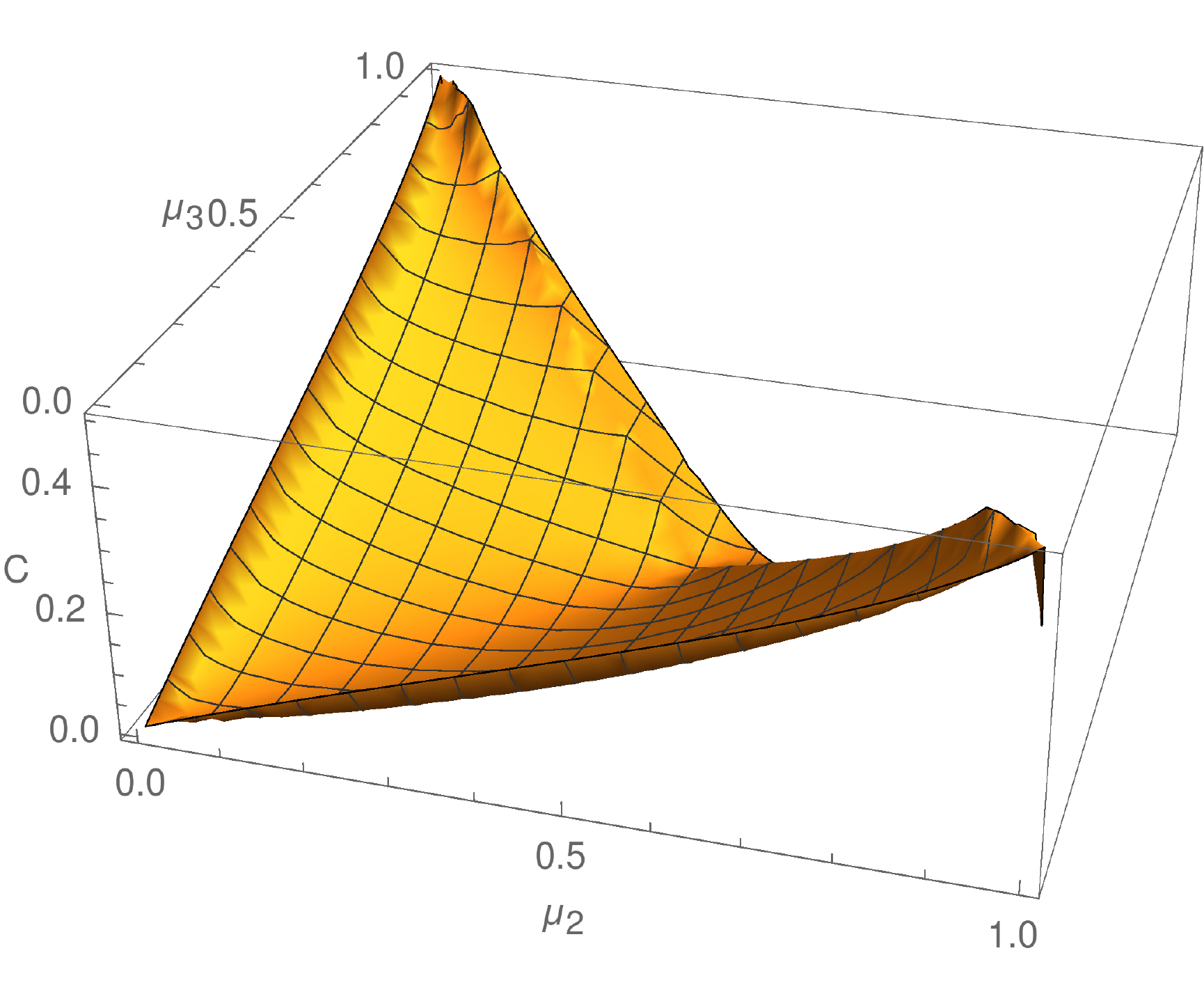}
\end{center}
\caption{(Color online) We plot $C$ obtained from the reduced two qubits density matrix  
 with $\mu_4=1$ as a function of $\mu_1$ and $\mu_2$  (a) and 
as a function of $\mu_2$ and $\mu_3$ (b).
plot (a) indicate that
the concurrence becomes maximum for $\mu_1=0$. Plot (b) suggests that 
$C$ becomes maximum when $\mu_2 \to 0$ with a finite $\mu_3$ and vice versa;$C$ is maximum for $\mu_2 \mu_3 \to 0$
except the case for ($\mu_2 \to 0$, $\mu_3$)  and ($\mu_3 \to 0$, $\mu_2$). The above features hold true for
any other value of $\mu_4 \neq 1$.} \label{fig8} \end{figure}

One can also show that concurrence, obtained  by taking the square root of the the above eigenvalues (\ref{eq:eigen2}) in appropriate order,
becomes maximum when $\mu_1 \to 0$ while $\mu_4$ and $\mu_2$ are non-zero. This is depicted in Fig.~(\ref{fig8}a) where 
concurrence is plotted as a function of $\mu_1$ and $\mu_2$; concurrence monotonically increases with decreasing $\mu_1$. This justifies the maximization of concurrence happens 
for $d=0$ case when $\mu_1 \to 0$. However, $\mu_2$ and $\mu_3$ are not always the same for $d=0$ case except $\delta \to 0$ limit. We consider
the simplified situation $\mu_2=\mu_3$ and $\mu_4=1$ to show explicitly that the concurrence indeed maximizes when $\mu_1 \to 0$ limit. This feature is 
common when $\mu_4 <1$ (for the $d \neq 0$ case) and $\mu_2 \neq \mu_3$ (for $d=0$ case).

Now we shall examine the 2nd peak observed in $C$ when $\mu_2 \mu_3 \to 0$ for any value of $d$. In order to investigate this
phenomena, we have to set $\mu_1=0$ in the reduced density matrix presented in Eq.~(\ref{eq:rho_s1}). An analytic closed form expression of 
eigenvalues in terms of $\mu_2$, $\mu_3$ and $\mu_4$ can not be obtained in this case even with simplified situation $\mu_4=1$. We present the concurrence, numerically obtained, in 
Fig.~(\ref{fig8}b) showing that the concurrence is almost a monotonic function of 
$\mu_2$ (while $\mu_3=0$) and $\mu_3$ (while $\mu_2=0$) except near the $\mu_2=1, \mu_3=0$ and $\mu_3=1, \mu_2=0$. Therefore, it is now clear that 
$C$ has a secondary maximum when $\mu_2\mu_3 \to 0$ except near the point $\mu_2= 0$ and $\mu_3=0$. This characteristics of 
concurrence is also seen for the case when $\mu_4\neq 1$ (for finite $d$).

Our observation suggests that $\mu_4$  always stays 
at unity for $d=0$ case but concurrence vanishes after the unentanglement time $t_{\rm UE}$. One can therefore infer that concurrence 
becomes independent of $\mu_4$ and vanishes subsequently when 
$\mu_1=\mu_2=0$; this can be easily seen by calculating  the concurrence from  the density matrix (\ref{eq:rho_s1}).
Now, for the case $d=0$, concurrence vanishes immediately after $\sqrt{\mu_3} \to 0$ as shown in Fig.~(\ref{fig6}a,b,c). Under the small
$\delta$ approximation, one can therefore conclude that
$\mu_2 \simeq \mu_3$ is the killing channel which can destroy the concurrence. On the other hand, for $d\neq0$, it is shown that concurrence vanishes 
after $\mu_3 \to 0$ instead of $\mu^{1/2}_3$ for $d=0$ case. We can say that in the $\delta \to 0$ limit $\mu_2$ or $\mu_3$ is the killing channel
for destroying the concurrence.
Once $\mu_1\to0$, $\mu_2\to0$ and $\mu_3\to0$ then concurrence becomes independent of the other channel $\mu_4$. This can be seen
in the temporal behavior of $C$ for $d=0$ and $d\neq 0$, where $C$ vanishes in presence of a finite $\mu_4$.

We shall now explain the existence of  a threshold time $t_{\rm TH}$ (see Fig.~(\ref{fig7}a,b,c)) in the light of above channel analysis.
 The  reduced density matrix of the two qubits up to the threshold time $t_{\rm TH}$ is given by
\be \rho_s(t) = {1\over 4}\left[ \begin{array}{cccc} 
1 & \sqrt{\mu_3} &\sqrt{\mu_3} & {\mu_3} \\
\sqrt{\mu_3} & 1 & {\mu_3} & \sqrt{\mu_3} \\
\sqrt{\mu_3} & {\mu_3} & 1 & \sqrt{\mu_3} \\
{\mu_3} & \sqrt{\mu_3} & \sqrt{\mu_3} & 1
\end{array} \right], \label{eq:rho_s2} \ee
this is obtained from (\ref{eq:rho_s1}) by considering in the numerically observed behavior of DCs
for $d\neq 0$ case, $\mu_4=\mu_1$ and $\mu_2=\mu_3$ and $\mu_3^2=\mu_1$.
One can compute the concurrence using the above density matrix (\ref{eq:rho_s2}). All the  eigenvalues 
of $\rho_s \hat\rho_s$ are same i.e., $\epsilon_1=\epsilon_2=\epsilon_3=\epsilon_4=(-1+\mu_3)^2$. This 
 yields  zero concurrence when $t<t_{TH}$ upto which the above density matrix (\ref{eq:rho_s2}) is valid.
 After the threshold time, $\mu_4 \neq \mu_1$, $\mu^2_3 \neq \mu_1$ and hence 
 concurrence becomes non-zero even if $\mu_2=\mu_3$. This threshold time increases with distance and becomes maximum when the 
 two qubits separated from each other maximally i.e., at $d=50$.

\begin{figure}[ht]
\begin{center}
\includegraphics[height=5.40cm,width=4.1cm]{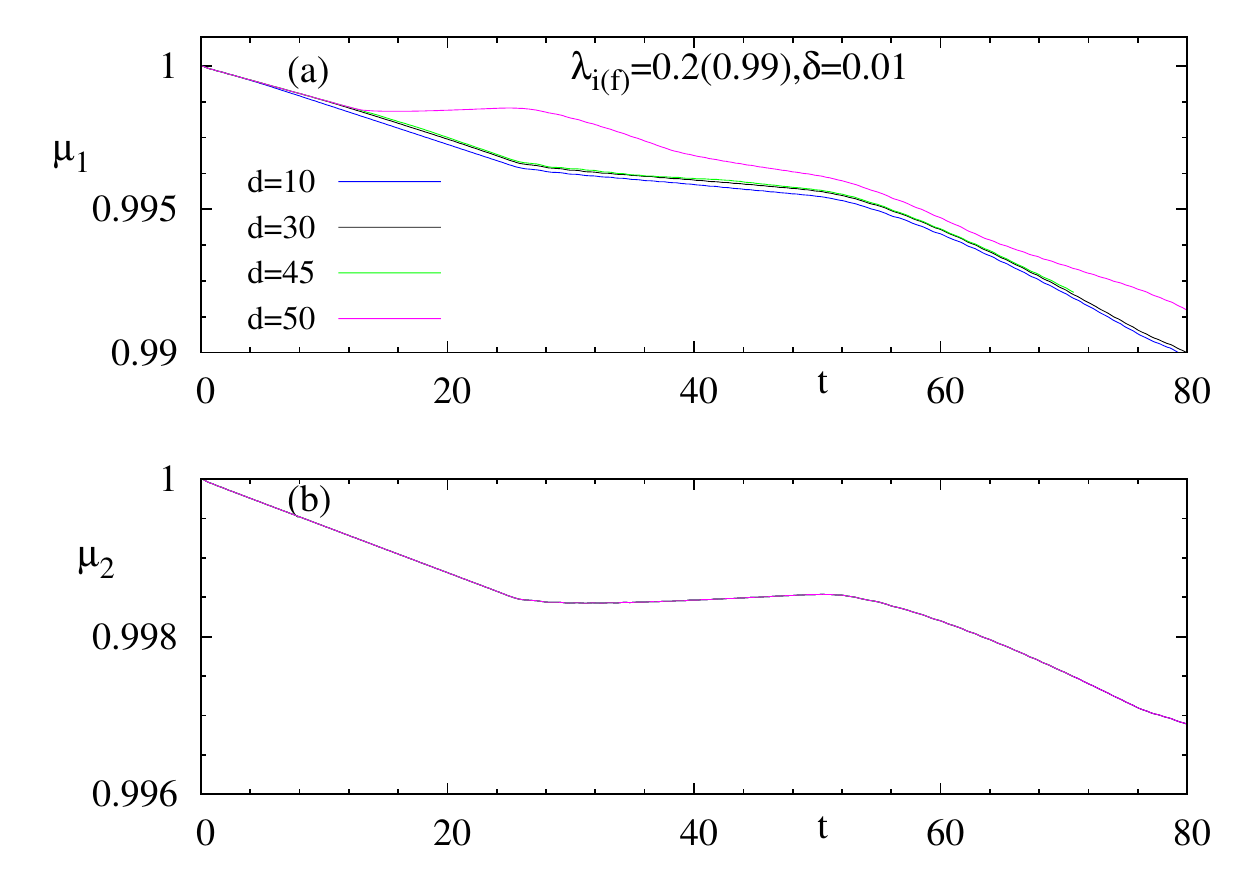}
\includegraphics[height=5.40cm,width=4.1cm]{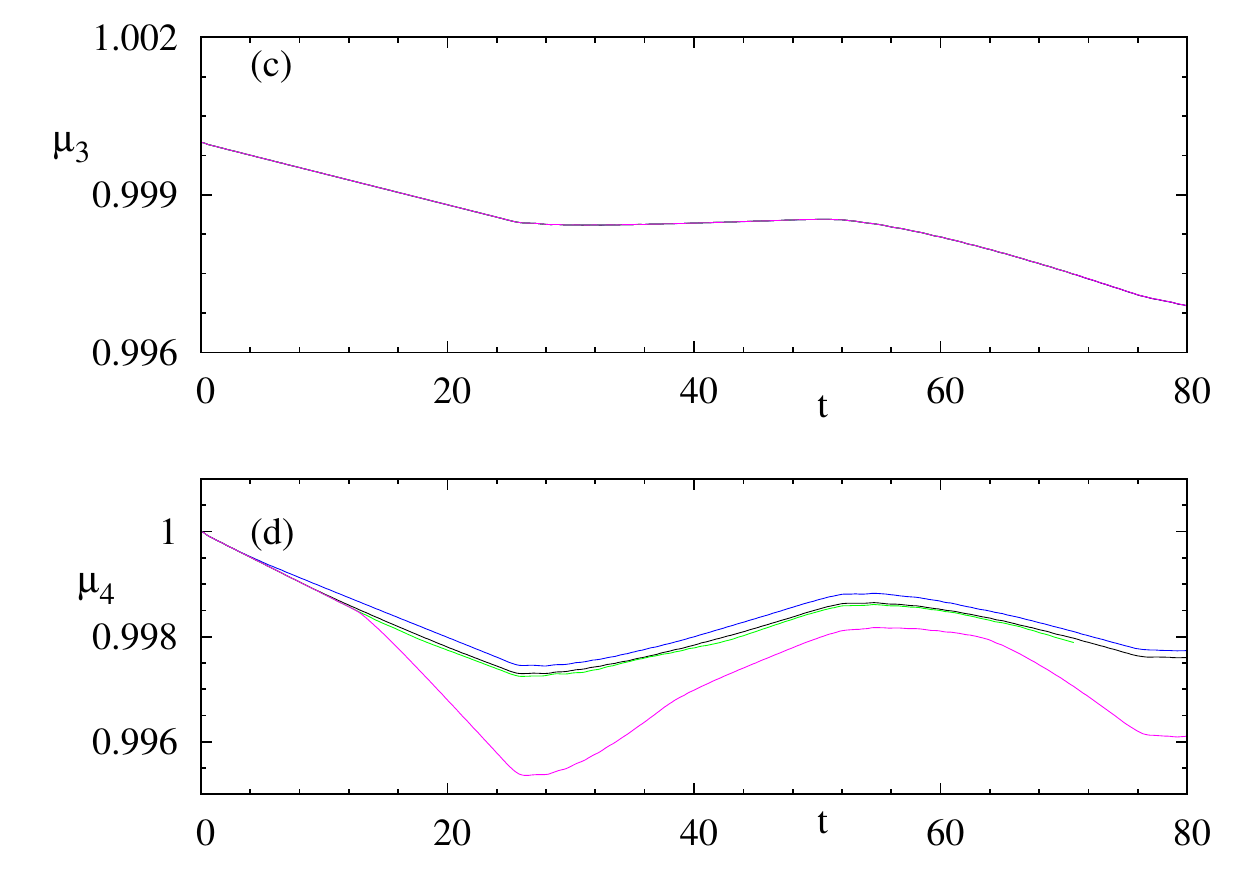}
\end{center}
\caption{ (Color online) The temporal behavior of four DCs  $\mu_1$ (a), $\mu_2$ (b), $\mu_3$ (c) and $\mu_4$ (d)
are plotted as a function of time choosing  different values of $d$.
Here, $N=100$, $\delta=0.1$ and $\la_{i,(f)}=0.2(0.99)$. The singular and the revival behavior are 
explained in the text.
} \label{fig9} \end{figure}

Furthermore, we explore the behavior of four DCs $\mu_1$, $\mu_2$, $\mu_3$ and $\mu_4$ as a function of time by varying the distance between the two 
qubits. Figure (\ref{fig9}a,d) show that the $\mu_1$ and $\mu_4$ channels are sensitive to $d$. When the two qubits are at 
symmetric position (i.e., $d=50$ for $N=100$ and PBC), both the channels 
exhibit  a singular behavior  at $t=t_S=N/(4v)$. This singular behavior at $t_S$ does not exist for non-symmetric situations. It is
also to be noted that additionally there is a revival time 
occurring at $t=t_R=N/(2v)$.
The other two channels $\mu_2$ and $\mu_3$ are absolutely insensitive to 
distance and as a result $t_S=N/(4v)$ is no longer  a special time scale for this channels even when $d=50$ (see Fig.~(\ref{fig9}b,c)).
For these above two channels echo exhibits the singular behavior at $t=t_S=N/(2v)$ which is twice of the earlier singular time 
scale for $\mu_1$ and $\mu_4$. These observation shall now be analyzed in the light of the quasi-particle picture.  

When the environmental spin chain is suddenly (and globally) quenched from the FM phase to 
the QCP, each of the environmental sites emit a pair of quasi-particles moving with opposite momentum
in opposite directions. Now, these two quasi-particles meet at $t=t_R=N/(2v)$ after traversing half of the environmental chain and  there is
a constructive interference causing a partial revival of the initial state (see Fig.~(\ref{fig10})).  The channels  $\mu_1$ and $\mu_4$ both involve two distinct Hamiltonians $(H_{\downarrow\downarrow}, H_{\uparrow\uparrow})$ and 
$(H_{\uparrow\downarrow}, H_{\downarrow\uparrow})$ which are different from each other in terms of the local transverse fields 
modified through the coupling of the qubits to the two sites, one can think of two extra separate
emitters, located at these two  sites  with  distance $d$ away from each other. Now, in the symmetric position 
$d=N/2$, quasi-particles need to travel only $d/2$ distance for such a revival to happen. Therefore,  $t_S=d/(2v)=N/(4v)$ and  $t_R= 2t_S$. (Over the
former time, two quasi-particles travel $N/4$ while in the latter they traverse a length of $N/2$.) In the case for $d\neq 50$, the singular time 
scale does not appear as no constructive interference of  two oppositely moving quasi-particles is not possible.

What is remarkable is that $\mu_2$ and $\mu_3$ do not exhibit the singular behavior at time  $t=t_S$; their dynamical evolution is only governed by the 
revival time scale $t_R$ (see Fig.~(\ref{fig9}b,d)). This is due to fact that one of the underlying Hamiltonians generates an extra pair
of quasi-particles
than that of the other Hamiltonian 
involved in  $\mu_2$ and $\mu_3$ (i.e., $H_{\downarrow\downarrow}$ is different from $H_{\uparrow\downarrow}$ by a
locally modified transverse field at a  single site and same for
$(H_{\uparrow\uparrow},H_{\uparrow\downarrow})$).   Therefore, the quasi-particle from this single emitter has to travel a distance $N/2$ to partially recover  the initial configuration even if  the qubits are separated by a distance $d=N/2$. Therefore, the number of independent emitters, originated from 
the structure of the two underlying Hamiltonian that governs the dynamics, dictates  the time scales of revival and singular behavior. These features are
observed in  the different channels of echo following a sudden quench.

\section{conclusion}
\label{conclude}

In this  paper,  we have studied a  GCSM where two qubits are locally  connected to the environmental transverse 
Ising spin chain in such a way that the local transverse field of the environment gets modified. Working in the weak
coupling limit, we explore the generation of the entanglement between the above pair of qubits, which are initially completely unentangled, both in equilibrium as well as non-equilibrium situations. In the former situation,   the concurrence between them is very small  in comparison to the non-equilibrium situation.
However, the role of quantum criticality manifest in the  behavior of concurrence  as it becomes maximum at the QCP and survives even when
$d$ is large; this behavior persists even in the non-equilibrium situation.   
Additionally, in the latter situation the  concurrence remains non-zero for longer time 
 if the quenching is within  the  same phase.  Furthermore, the time at which concurrence exhibits a primary peak is 
 inversely proportional to the coupling  strength.  Remarkably,  we show that there exists a threshold time above which concurrence 
 becomes finite for $d \neq 0$.

Analyzing the behavior of concurrence using different DCs,  we show that the number of effectively independent channels in the  is determined by the distance $d$ 
and the  local difference of the transverse field of the two underlying Hamiltonians governing the time evolution of the channels. This is a universal observation
that holds true for both  the equilibrium and non-equilibrium situations. The decoherence is maximum in the channel
($\mu_1$) that involves the maximum difference in 
the local transverse fields of two Hamiltonians dictating the time evolution; consequently, the short time decay of the echo is most rapid for this
channel in all the situation discussed in the paper.  Hence, the time at which  this most rapidly decaying DC  decays
to zero, there is  the  primary peak in the concurrence. On the other hand,  the product of the intermediate decaying channels ($\mu_2 \mu_3$) is responsible for the secondary peak of the concurrence. 
Besides these above two common features, there exists a markedly 
different connection of concurrence with these DCs for a finite separation $d$. The subsequent temporal decay of concurrence is attributed
to the behavior of the killing channel which is an 
intermediate decaying channel ($\mu_3$); for $d=0$, the concurrence decays to 
zero when $\sqrt{\mu_3}\to 0$,  while the condition gets modified to $\mu_3 \to 0$,  for $d \neq 0$. 
Additionally, we analyze  the existence of a finite threshold time using the channel analysis. 
Comparing the behavior of the  concurrence in the non-equilibrium case with the DP obtained from the reduced single qubit 
density matrix,  we explain the interesting observation  that the dephasing rate is always higher than 
the unentanglement rate. This  is a consequence of  the fact that DCs appear additively in DP and additionally, the most rapidly DC is absent in the $2\times 2$
reduced density matrix.
Finally, we make resort to quasi-particle picture to characterize the temporal evolution of different channels, following a critical quench,
with different values of the distance between the qubits as a function of time. We explain the singular and revival behavior of different channels analyzing the
interference  of quasi-particles. Remarkably, the singular behavior in the evolution shows up only for the maximally  and minimally 
decaying channel denoted by $\mu_1$ and  $\mu_4$, respectively.

\section{Acknowledgement}
TN and AD thank Uma Divakaran for her valuable suggestions and numerical help. AD Acknowledges SERB, DST for financial
support.


\begin{thebibliography}{30}

\bi{einstein35} A. Einstein, B. Podolsky and N. Rosen, Phys. Rev. {\bf 47}, 777 (1935).

\bi{werner89} R. F. Werner, Phys. Rev. A {\bf 40}, 4277 (1989).

\bi{bennett96} C. H. Bennett, D. P. DiVincenzo, J. A. Smolin and W. K. Wootters, Phys. Rev. A {\bf 54}, 3824 (1996).


\bi{nielsen00} M. A. Nielsen and I. L. Chuang, {\it Quantum Computation and
Quantum Information}(Cambridge University Press, Cambridge, UK, 2000).

\bi{vedral07} V. Vedral, {\it Introduction to Quantum Information Science}(Oxford University Press, Oxford, UK, 2007).


\bi{chakrabarti96} B. K. Chakrabarti, A. Dutta and P. Sen, {\it Quantum Ising Phases and 
transitions in transverse Ising Models}, m41 (Springer, Heidelberg,1996).

\bibitem{sachdev99} S. Sachdev, 
{\it Quantum Phase Transitions}(Cambridge University Press, Cambridge, England,1999). 

\bi{polkovnikov11} A. Polkovnikov, K. Sengupta, A. Silva and M. Vengalattore, Rev. of Mod. Phys. {\bf 83}, 863 (2011).




\bi{dutta15} A. Dutta, G. Aeppli, B. K. Chakrabarti, U. Divakaran, T. 
Rosenbaum and D. Sen, \textit{Quantum Phase Transitions in Transverse Field 
Spin Models: From Statistical Physics to Quantum Information} (Cambridge 
University Press, Cambridge, 2015).
\bi{osterloh02} A. Osterloh, L. Amico, G. Falci and R. Fazio, Nature {\bf 416}, 608 (2002).

\bi{osborne02} T. J. Osborne and M. A. Nielsen, Phys. Revs. A, {\bf 66}, 032110 (2002).

\bi{vidal03} G. Vidal, J. I. Latorre, E. Rico and A. Kitaev, Phys. Rev. Lett. {\bf 90}, 227902 (2003).

\bi{wu04} L. A. Wu, M. S. Sarandy and D. A. Lidar, Phys. Rev. Lett, {\bf 93}, 250404 (2004).

\bi{chen06} Y. Chen, P. Zanardi, Z.D. Wang, and F. C. Zhang, New. J. Phys. {\bf 8}, 97 (2006).

\bi{gu04} S. J. Gu, S. S. Deng, Y. Q. Li and H. Q. Lin, Phys. Rev. Lett. {\bf 93}, 086402 (2004).


\bi{amico08} L. Amico, R. Fazio, A. Osterloh, V. Vedral, Rev. Mod. Phys. {\bf 80}, 517-576 (2008).
\bi{wootters01} W. K. Wootters, {\it Qunatum Inf. Comput.} {\bf I}, 27 (2001).

\bi{horodecki01} M. Horodecki, {\it Qunatum Inf. Comput.} {\bf I}, 3 (2001).

\bi{horodecki09} R. Horodecki, P. Horodecki, M. Horodecki and K. Horodecki, Rev. Mod. Phys. {\bf 81}, 865 (2009).

\bi{ollivier01} H. Ollivier and W. H. Zurek, Phys. Rev. Lett. {\bf 88}, 017901 (2001); ; W. H. Zurek, Rev. Mod. Phys. {\bf 75}, 715 (2003).

\bi{luo08} S. Luo, Phys. Rev. A {\bf 77}, 042303 (2008). 

\bi{sarandy09} M. S. Sarandy, Phys. Rev. A, {\bf 80}, 022108 (2009).
\bi{calabrese05} P. Calabrese and J. Cardy, J. Stat. Mech: Theor. Exp. (2005) {\bf P04010}; P. Calabrese and J. Cardy, J. Stat. Mech: Theor. Exp. (2007) {\bf P06008};
L. Cincio, J. Dziarmaga, M. M. Rams and W. H. Zurek, Phys. Rev. A {\bf 75}, 052321 (2007).
\bi{joos03} E. Joos, H. D. Zeh, C. Keifer, D. Giulliani, J. Kupsch and
I. -O. Stamatescu, \textit{Decoherence and appearance of a classical world in a quantum theory}, (Springer Press, Berlin) (2003).
 
\bi{plenio99} M. B. Plenio, S. F. Huelga, A. Beige and P. L. knight, Phys. Rev. A {\bf 59}, 2468 (1999);  S. Bose,
P. L. Knight, M. B. Plenio and V. Vedral, Phys. Rev. Lett. {\bf 83}, 5158 (1999); P. Horodecki, Phys. Rev. A {\bf 63}, 022108 (2001);
X. X. Yi, C. S. Yu, L. Zhou and H. S. Song, Phys. Rev. A {\bf 68}, 052304 (2003); S. Shresta, C. Anastopoulos, A. Dragulescu and B. L.
Hu, Phys. Rev. A {\bf 71}, 022109 (2005).

\bi{viola98} L. Viola and S. lioyd, Phys. Rev. A {\bf 58}, 2733 (1998).

\bi{rossini08} D. Rossini, T. Calarco, V. Giovannetti, S. Montangero and R. Fazio, Phys. Rev. A {\bf 75}, 032333 (2007); 
D. Rossini, P. Facchi, R. Fazio, G. Florio, D. A. Lidar, S. Pascazio, F. Plastina and P. Zanardi, Phys. Rev. A {\bf 77},
052112 (2008); T. Fogarty, E. Kajari, B. G. Taketani, A. Wolf, T. Busch and G. Morigi, Phys. Rev. A {\bf 87}, 050304 (2013); B. G. taketani,
T. Fogarty, E. kajari, T. Busch and G. Morigi, Phys. Rev. A {\bf 90}, 012312 (2014).

\bi{cucchietti05} F. M. Cucchietti, J. P. Paz and W. H. Zurek, Phys. Rev. A {\bf 72}, 052113 (2005); F. M. Cucchietti, 
S. Fernandez-Vidal, J. P. Paz, Phys. Rev. A {\bf 75}, 032337 (2007); Z. -G. Yuan, P. Zhang and S. -S. Li, Phys. Rev. A {\bf 76}
042118 (2007).

\bi{quan06} H. T. Quan, Z. Song, X. F. Liu, P. Zanardi, and C. P. Sun, Phys. Rev. Lett. {\bf 96}, 140604 (2006).

\bi{sharma12} S. Sharma, V. Mukherjee and A. Dutta, Eur. Phys. J B {\bf 85}, 143 (2012).


\bi{damski11a} B. Damski, H. T. Quan, W. H. Zurek, Phys. Rev. A {\bf 83}, 062104 (2011).

\bi{nag12a} T. Nag, A. Dutta and A. Patra, Int. J. Mod. Phys. B {\bf 27}, 1345036, (2013).
\bi{nag12b}T. Nag, U. Divakaran and A. dutta, Phys. Rev. B, {\bf 86}  020401 (R) (2012).

\bi{roy13} S. Roy, T. Nag and A. Dutta, EPJ B {\bf 86}, 204 (2013).

\bi{patel13} A. A. Patel, S. Sharma and A. Dutta, EPJ B {\bf 86}, 367 (2013).

\bi{mukherjee12} V. Mukherjee, S. Sharma and A. Dutta, Rev. B, {\bf 86}  020301 (R) (2012).

\bi{sharma14} S. Sharma, A. Russomanno, G. E. Santoro and A. Dutta, Euro. Phys. Lett., {\bf 106} 67003 (2014). 

\bi{suzuki15} S. Suzuki, T. Nag and A. Dutta, Phys. Rev. A, {\bf 93}, 012112(2016).

\bi{nag15} T. Nag, arXiv: 1512.02959 (2015).
\bi{sengupta09} K. Sengupta and D. Sen, Phys. Rev. A {\bf 80}, 032304 (2009).

\bi{nag11} T. Nag, A. Patra and A. Dutta, Jour. Stat. Mech. (2011) {P08026}.

\bi{zurek05} W. H. Zurek, U. Dorner and P. Zoller, Phys. Rev. Lett.
{\bf 95}, 105701 (2005)

\bi{polkovnikov05} A. Polkovnikov, Phys. Rev. B {\bf 72}, 161201 (2005).

\bi{yuan07} Z. G. Yuan, P. Zhang and S. -S. Li, Phys. Rev. A {\bf 76}, 042118 (2007).
\bi{sun07} Z. Sun, X. Wang and C. P. Sun, Phys. Rev. A {\bf 75}, 062312 (2007).
\bi{liu10} B. -Q. Liu, B. Shao and J. Zou, Phys. Rev. A {\bf 82}, 062119 (2010).


\bi{yi06} X. X. Yi, H. T. Cui and L. C. Wang, Phys. Rev. A {\bf 74}, 054102 (2006).
\bi{ai08} Q. Ai, T. Shi, G. Long and C. P. Sun, Phys. Rev. A {\bf 78}, 022327 (2008).
\bi{greiner02} M. Greiner {\it etal}, Nature {\bf 415} 39 (2002).
\bi{duan03} L. M. Duan, E. Demler and M. D. Lukin, Phys. Rev. Lett. {\bf 91}, 090402 (2003); D. Porras and J. I. Cirac, Phys. Rev. Lett.
{\bf 92}, 207901 (2004); J. K. Pachos and M. B. Plenio, Phys. Rev. Lett. {\bf 93}, 056402 (2004).

\bibitem{zhang09} J. Zhang, F. M. Cucchietti, C. M. Chandrashekar, M. Laforest, C. A. Ryan, M. Ditty, 
A. Hubbard, J. K. Gamble, and R. Laflamme, Phys. Rev. A {\bf 79}, 012305 (2009).		


\bi{cormick08} C. Cormick and J. P. Paz, Phys. Rev. A {\bf 78}, 012357 (2008).

\bi{wendenbaum14} P. Wendenbaum, B. Taketani and D. Karevski, Phys. Rev. A {\bf 90}, 022125 (2014); C. Cormick and J. P. Paz, Phys.
Rev. A {\bf 78}, 012357 (2008). 

\bi{hartmann06} M. J. Hartmann, M. E. Reuter and M. B. Plenio, New J. Phys. {\bf 8}, 94 (2006).

 
 \bi{lieb61}E. Lieb and T. Schultz and D. Mattis, Ann. Phys.,NY {\bf 16}, 37004 (1961).

\bi{pfeuty70}P. Pfeuty, Ann. Phys. (NY) {\bf 57}, 79 (1970).

\bi{keyl10} M. Keyl and D. -M. Schlingemann, J. Math. Phys. {\bf 51}, 023522 (2010).
\bi{yu03} T.  Yu and J. H. Eberly, Phys. Rev.  B {\bf 68}  165322   (2003); K. Ann, G. Jaeger, Phys. Rev. B {\bf 75} 115307 (2007).


 
\end{thebibliography}
\end{document}